\documentclass[12pt]{article}

\usepackage{scicite,graphicx,amsmath,amssymb,bm,epstopdf,latexsym,hyperref,braket,mathtools}
\usepackage[normalem]{ulem} 

\newenvironment{sciabstract}{%
\begin{quote} \bf}
{\end{quote}}

\title{Hamiltonian engineering of spin-orbit coupled fermions in a Wannier-Stark optical lattice clock}

\author
{Alexander Aeppli$^{1, \dagger}$, Anjun Chu$^{1,2, \dagger}$, Tobias Bothwell$^{1, \dagger}$, Colin J. Kennedy$^{1}$,\\
Dhruv Kedar$^{1}$, Peiru He$^{1,2}$, Ana Maria Rey$^{1,2\ast}$, Jun Ye$^{1\ast}$
\\
\normalsize{$^{1}$
JILA, National Institute of Standards and Technology,}\\
\normalsize{and Department of Physics, University of Colorado, Boulder, CO 80309}
\\
\normalsize{$^{2}$ Center for Theory of Quantum Matter, University of Colorado, Boulder, CO 80309}
\\
\normalsize{$^\ast$Corresponding authors; e-mail: ye@jila.colorado.edu; arey@jilau1.colorado.edu}\\
\normalsize{$^\dagger$These authors contributed equally to this work}
}

\date{}

\begin{document} 




\maketitle

\begin{sciabstract}
Engineering a Hamiltonian system with tunable interactions provides opportunities to optimize performance for quantum sensing and explore emerging phenomena of many-body systems. 
An optical lattice clock based on partially delocalized Wannier-Stark states in a gravity-tilted shallow lattice supports superior quantum coherence and adjustable interactions via spin-orbit coupling, thus presenting a powerful spin model realization. The relative strength of the on-site and off-site interactions can be tuned to achieve a zero density shift at a `magic' lattice depth. This mechanism, together with a large number of atoms, enables the demonstration of the most stable atomic clock while minimizing a key systematic uncertainty related to atomic density. Interactions can also be maximized by driving off-site Wannier-Stark transitions, realizing a ferromagnetic to paramagnetic dynamical phase transition.
\end{sciabstract}

The joint advance of quantum metrology and quantum simulation provides exciting new opportunities to explore the frontiers of measurement science and the emergence of many-body complexity. An outstanding example has been the development of optical lattice clocks (OLCs) where excellent quantum coherence and exquisite quantum control of many atoms have enabled rapid advances in metrological capabilities~\cite{Campbell2017,Oelker2019, Nicholson2015,ludlow2015,mcgrew2018,zheng2021high}, culminating in the recent demonstration of clock measurement precision at $7 \times 10^{-21}$ and near minute-long atomic coherence~\cite{bothwell2021}. To achieve this level of performance, we use a shallow, vertically aligned optical lattice.
The acceleration due to local gravity lifts the degeneracy of neighboring sites, supporting partially delocalized Wannier-Stark eigenstates. This trapping scheme, first suggested in 2005~\cite{lemonde2005}, allows us to operate the clock at substantially smaller lattice depths, greatly suppressing detrimental motional, light scattering, and atomic density induced decoherence.

The use of tilted optical lattices to manipulate motional degrees of freedom in ultracold gases has been widely reported. They have been used to suppress direct tunneling but not spin transport and realize new types of spin Hamiltonians\cite{Dimitrova2020, Brown2015,Trotzky2008}, generate spin-orbit coupling via laser-assisted tunneling~\cite{Miyake2013,Aidelsburger2013,Aidelsburger2015,Kennedy2015},  emulate magnetic models in spinless bosons~\cite{Simon2011,Meinert2013}, probe non-ergodicity due to kinetic constraints~\cite{Scherg2021} and subdiffusive transport~\cite{Guardado2020} in Fermi-Hubbard chains and many-body localization in trapped ions~\cite{morong2021}, as well as measure gravity in Raman interferometers \cite{Tino2021,Xu2019}. In this work, we demonstrate how a tilted optical lattice combined with pristine quantum coherence and exquisite spectral resolution offer new capabilities to engineer, drive, and understand many-body systems.

As we continue to push the OLC to new levels of precision, a key remaining issue for clock accuracy is related to frequency shifts associated with atomic interactions. Quantum statistics dictates that identical fermions experience only odd partial wave interactions that are suppressed at ultralow temperatures \cite{campbell2009,ludlow2011,swallows2011,lemke2011,martin2013,rey2014}. Yet, even the weak elastic and inelastic $p$-wave collisions were found to significantly affect clock operation and limit the number of interrogated atoms at deep lattice depths. As atoms delocalize along neighboring sites in the shallow lattice, $p$-wave collisions are reduced but $s$-wave interactions can emerge from the spin-orbit coupling (SOC) generated by the differential clock laser phase\cite{kolkowitz2016,bromley2018,wall2016}. The superior quantum coherence obtained in our gravity-tilted optical lattice clock stems from better control over motional and internal degrees of freedom \cite{bothwell2021}, allowing the engineering of $s$- and $p$-wave interactions in driven spin-orbit coupled fermionic atoms.  By operating at the `magic' lattice depth where $s$-wave interactions precisely cancel residual $p$-wave interactions, we reduce atomic-interaction induced shifts in our 1D lattice clock to a fractional frequency shift of $5.0(1.7) \times 10^{-21}$ per atom at a single site.

We further explore the tunability of atomic interactions by driving a site-changing Wannier-Stark transition. This leads to an atomic superposition that not only carries a distinct internal label but also features different motional orbitals. As a consequence, $s$-wave interactions are significantly enhanced. This gives rise to a many-body dynamical phase transition between dynamical ferromagnetic and paramagnetic states controlled by the interplay between the clock drive and atomic interactions. Although similar dynamical phase transitions have been observed in trapped ions~\cite{Zhang2017}, superconducting qubits\cite{Xu2020}, and atoms in cavities~\cite{Muniz2020} and optical traps~\cite{Chu2020}, here we use in situ imaging to locally resolve the emergence of a non-linear excitation lineshape as a function of atom number.

\begin{figure}[htp!]
\centerline{
\includegraphics[width=12 cm]{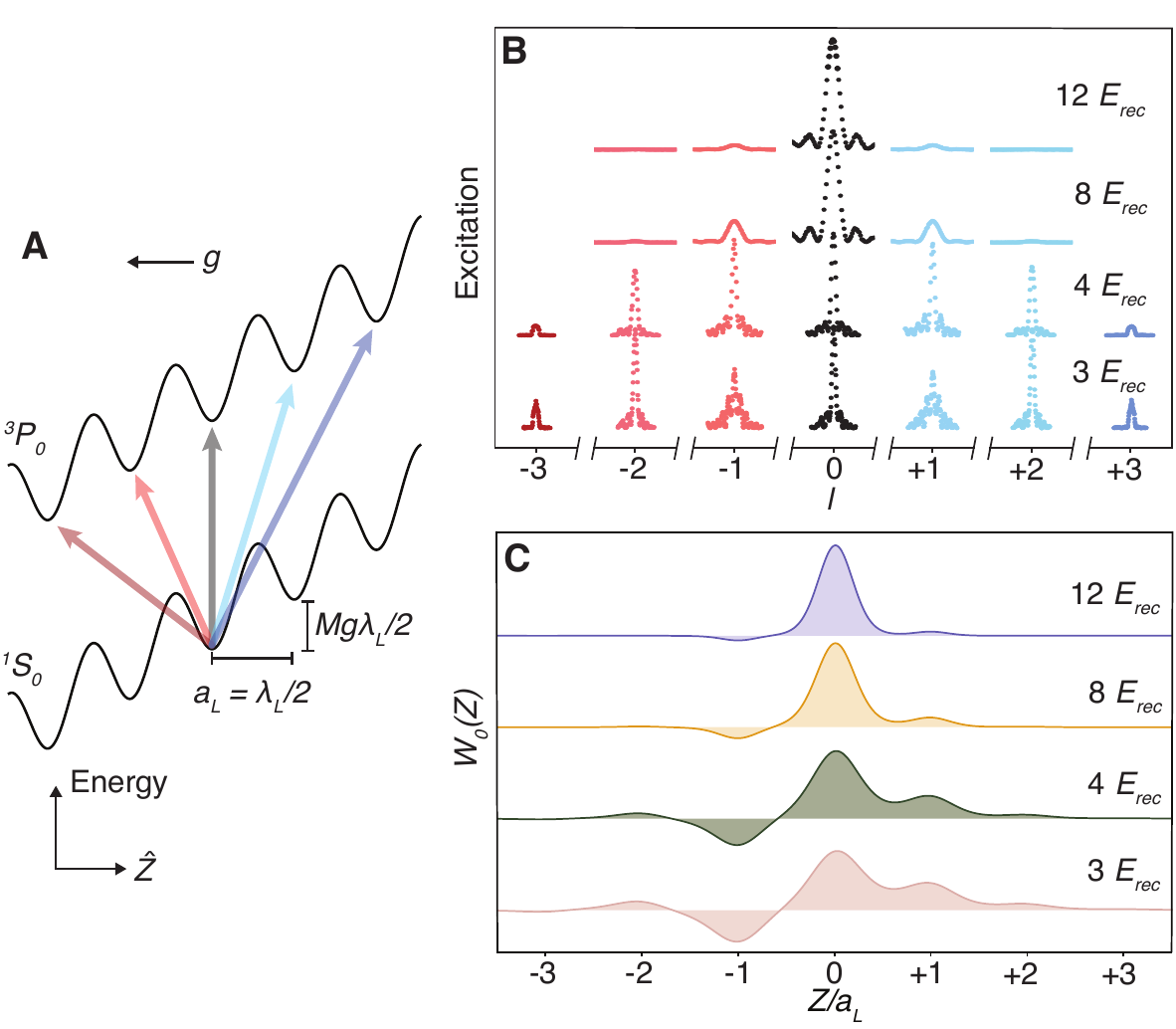} }
\caption{\textbf{The Wannier-Stark Clock.} (\textbf{A}) We trap $^{87}$Sr atoms in a $1D$ optical lattice along the $\hat{Z}$ direction aligned with local gravitational acceleration $g$. 
This type of external confinement realizes Wannier-Stark states, eigenstates of the joint lattice plus gravitational potential. 
The $n^{\text{th}}$ Wannier state $W_n (Z)$ is centered at lattice site $n$ and has energy $M g a_L n$, where $M$ is the mass of $^{87}$Sr and $a_L =\lambda_L/2$ is the lattice spacing with lattice wavelength $\lambda_L$. 
The Wannier-Stark ladder creates a set of transitions from the ground ($| g \rangle\equiv |{}^1S_0,m_F=\pm 5/2\rangle$) to clock ($| e \rangle\equiv |{}^3P_0,m_F=\pm 3/2\rangle$) state at different lattice sites accessible by the differential clock laser phase between them.
The black line indicates a carrier $|g \, ; \, W_n\rangle \rightarrow |e\,;\,W_{n} \rangle$ transition.
At shallow lattice depths, a set of off-site transitions $|g \,;\, W_n\rangle \rightarrow |e \,;\, W_{n \pm l} \rangle$ for integer $l$ are indicated by blue and red lines.
(\textbf{B}) At shallow lattice depths, the atomic wavefunction becomes delocalized, allowing $|g \,;\, W_n\rangle \rightarrow |e \,;\, W_{n \pm l} \rangle$ transition for a range of $l$ to be addressed. 
Here we show Rabi scans of these transitions at four different lattice depths, given in lattice photon recoil energy ($E_{rec}$).
For each lattice depth, the pulse area was adjusted for a $\pi$ pulse on the carrier transition.
(\textbf{C}) The wavefunction $W_0(Z)$ for the four corresponding lattice depths, illustrating the tunable delocalization due to the interplay between lattice and gravitational potential.}
\label{fig_1}
\end{figure}

\paragraph*{System in Consideration}
Several hundred thousand nuclear-spin-polarized fermionic $^{87}$Sr atoms are cooled via standard techniques and loaded into a vertical one-dimensional optical lattice that defines the $\hat{Z}$ axis \cite{bothwell2021}. 
We load the lattice at a depth of $300$ lattice photon recoil energies ($E_{\rm rec}$) at $800$ nK. We employ sideband cooling to prepare the sample in the lowest motional band along the $\hat{Z}$ axis. Perpendicular to the lattice axis, the atoms are weakly confined and thermally populate the resultant radial modes. 
We then adiabatically reduce the lattice depth to a much lower operational depth with a correspondingly reduced radial temperature measured with Doppler spectroscopy (see SOM). 

The gravitational potential with local acceleration $g$ adds a linear energy gradient across the lattice, with the combined single-particle Hamiltonian supporting Wannier-Stark (WS) eigenstates.
The WS state $ W_n (Z)$ is centered at lattice site $n$ and has eigenenergy $M g a_L n$, where $M$ is the mass of $^{87}$Sr and $a_L =\lambda_L/2$ is the lattice site spacing (Fig.~\ref{fig_1}A). Here, we use the strontium `magic' wavelength $\lambda_L = 813$ nm, guaranteeing identical confinement for both clock states. 

The clock laser $\lambda_c = 698$ nm, aligned along the lattice, drives the ultranarrow $|{}^1 S_0,m_F=\pm5/2\rangle$ $\rightarrow |{}^3P_0,m_F=\pm3/2\rangle$ ($|g \rangle \rightarrow |e \rangle$) clock transition, where $m_F$ is the nuclear Zeeman level. This $\sigma$-polarized transition is the least magnetically sensitive clock transition in $^{87}$Sr.
Because the clock laser wavelength differs from the lattice spacing, adjacent lattice sites see a different clock phase $ \varphi  = \pi\lambda_L/\lambda_c \approx 7\pi  /6$. 
This phase difference generates SOC when the lattice depth is sufficiently low  for  atoms to tunnel during the course of the experiment. Thus, when tuned to appropriate frequencies, the clock laser effectively couples Wannier-Stark states between different lattice sites, i.e. $|g \,; \, W_n\rangle \rightarrow |e \,; \, W_{n+l} \rangle$, for a range of integer $l$. The corresponding Rabi frequency $\Omega_l$ set by the wavefunction overlap is
\begin{equation}
\Omega_l \propto \exp\left(-\frac{\lambda_L^2}{4\lambda_c^2\sqrt{V_0}}\right) \mathcal{J}_l \left( \frac{4 J_0}{M g a_L} \sin ( \varphi/2) \right).
\label{omega}
\end{equation}
Here, $\mathcal{J}_l$ is a Bessel function, $J_0$ is the nearest neighbor tunneling energy of the ground band, and $V_0$ is the lattice depth in $E_{rec}$. 

We utilize Rabi spectroscopy in a dilute ensemble to  demonstrate the partially delocalized nature of the single-particle wavefunctions in shallow, tilted lattices of four different values of $V_0$, shown in Fig.~\ref{fig_1}B.  The corresponding WS wavefunctions $W_0(Z)$ are shown in Fig.~\ref{fig_1}C. For each $V_0$, we optimize the transition probability on the carrier transition, $|g \,; \, W_n\rangle \rightarrow |e \,; \, W_n \rangle$. For $V_0 = 12~E_{rec}$, the atoms are still well localized, and thus the $|g \, ; \, W_n\rangle \rightarrow |e \, ; \, W_{n\pm 1} \rangle$ transition amplitudes are significantly suppressed in comparison to the carrier.  As $V_0$ is reduced, we resolve a set of Rabi lines spectrally separated by $Mga_L/h=867$ Hz, where $h$ is Planck's constant. 
At $4~E_{rec}$, the Rabi frequency for the carrier and $|g \, ; \, W_n\rangle \rightarrow |e \, ; \, W_{n\pm 1} \rangle$ transitions are roughly equivalent.
At $3~E_{rec}$, the carrier and $|g \, ; \, W_n\rangle \rightarrow |e \, ; \, W_{n\pm 2} \rangle$ have similar Rabi frequencies, while the $|g \, ; \, W_n\rangle \rightarrow |e \, ; \, W_{n\pm 1} \rangle$ transition has the greatest Rabi frequency and is thus overdriven.
At low atomic density,  we observe  coherence times well past $10$ s on $|g \,;\, W_n\rangle \rightarrow |e \,;\, W_{n + 1} \rangle$ (see SOM). 

\paragraph*{Theoretical Model}
Under our operating conditions, where the collisional rate for motional relaxation is smaller than the internal spin dynamics and trap frequencies, atoms remain effectively frozen in single-particle eigenstates during clock interrogation.
Since all atoms are initially prepared in a single internal state, Fermi statistics forbids double occupancy of motional states. Under these conditions, the quantum dynamics can be described with a spin Hamiltonian in energy space spanned by the appropriate single-particle trap eigenmodes~\cite{martin2013,rey2014,zhang2014spectroscopic,smale2019,Chu2020}. We identify a two level system for an atom in mode $\bf{n}$ as $|\uparrow_{\bf{n}}\rangle\equiv|e;n_X,n_Y,W_{n}\rangle$ and $|\downarrow_{\bf{n}}\rangle\equiv|g;n_X,n_Y,W_{n}\rangle$. Here, $n_X$ and $n_Y$ label the radial harmonic oscillator modes.

Two dominant types of interatomic interactions determine the coupling constants in the spin model: local interactions between atoms within a single lattice site and nearest-neighbour interactions between atoms in adjacent sites.  Next to nearest-neighbour interactions are typically small for the operating conditions in our system and are neglected. The couplings between radial harmonic oscillator modes are highly collective as shown in prior experiments \cite{martin2013,zhang2014spectroscopic,rey2014}. Therefore, to an excellent approximation, we define collective spin operators at each lattice site  after summing  over occupied harmonic oscillator modes, $\hat{S}^{x,y,z}_n=\sum_{n_X,n_Y}\hat{S}_{\bf n}^{x,y,z}$.   The dynamics of the collective spin vector $\langle \hat{\mathbf{S}}_n\rangle=\{\langle \hat{S}^x_n\rangle,\langle \hat{S}^y_n\rangle,\langle \hat{S}^z_n\rangle\}$  is  described by the following mean-field equation of motion
written in a gauge frame where the laser drive is homogeneous (see SOM):

\begin{equation}
    \frac{\mathrm{d}}{\mathrm{d}t}\langle\hat{\mathbf{S}}_n\rangle=\mathbf{B}^{\perp}\times\langle\hat{\mathbf{S}}_n\rangle.
    \label{MFS}
\end{equation} 
The synthetic magnetic field $\mathbf{B}^{\perp}$ contains contributions of the laser drive with detuning $\delta$ from the bare transition and the self-generated interactions terms:
\begin{equation}
    \mathbf{B}^{\perp}=\{\Omega_0,0,-\delta+2(\chi_0+\chi_1)\langle \hat{S}^z\rangle+C_0N_{\mathrm{loc}}\}.
    \label{eq:b2}
\end{equation} 
Here $\langle \hat{S}^z\rangle= \frac{1}{2L+1}\sum_{m=-L}^{L} \langle \hat{S}^z_{n+m}\rangle$ is the average magnetization over a region of  $2L+1\sim 15$ lattice sites (corresponding to 1 camera pixel or $6$ $\mu$m in our imaging spectroscopy) centered around $n$. $N_{\mathrm{loc}}$ is the number of atoms per lattice site averaged over the same region. The couplings, $\chi_0=\eta_0(V_{ee}+V_{gg}-2V_{eg})/2, C_0=\eta_0(V_{ee}-V_{gg})/2$, and $\chi_1=-\eta_1 U_{eg}(1-\cos\varphi)$, respectively describe thermally averaged $p$-wave and $s$-wave interaction parameters between internal clock states, as well as on the on-site ($\eta_{0}$) and nearest-neighbour ($\eta_{1}$)  overlap  matrix elements along the lattice. 
In the absence of SOC, $\varphi=0$, the $s$-wave interactions vanish. 

Without interactions, the collective spin  features  a characteristic Rabi lineshape profile when driven during  a pulse area $\Omega_0 T=\pi $ with excitation fraction $n_\uparrow(t)= \langle \hat{S}^z(t)\rangle/ N_{\mathrm{loc}} +1/2$, symmetric and centered around $\delta=0$. With interactions the time evolution takes place in the presence of an additional self-generated axial  magnetic field-like term that induces  a non-linear response, resulting in an asymmetric lineshape. 
A simple estimation of the density shift can be obtained by setting it to be the value of $\delta$ at which $\mathbf{B}^{\perp}_z=0$:
\begin{eqnarray}
&&\Delta\nu_{\alpha\to\beta}=\Delta\nu_{\alpha\to\beta}^s+\Delta\nu_{\alpha\to\beta}^p, \label{DS}\\
&&    2\pi\Delta\nu_{\alpha\to\beta}^p\approx2\chi_0\varsigma^z_{\alpha\to\beta}+C_0, \quad \quad  2\pi\Delta\nu_{\alpha\to\beta}^s\approx 2\chi_1\varsigma^z_{\alpha\to\beta}.
\end{eqnarray} 
Here, $\Delta\nu_{\alpha\to\beta}^{s,p}$ are the $s$-wave and $p$-wave contributions to the density shift, $\alpha$ and $\beta$ indicate initial and final states, $|g \rangle$ or $|e \rangle$. 
$\varsigma^z_{\alpha\to\beta}$ is a fitting parameter that accounts for the time evolution of $\langle \hat{S}^z\rangle/N_{\mathrm{loc}}$ during the Rabi dynamics, which depends on the details of the Rabi drive such as the pulse area, excitation fraction, and initial conditions used in the experiment (see SOM).
\begin{figure}[htp!]
\centering
\includegraphics[width=12 cm]{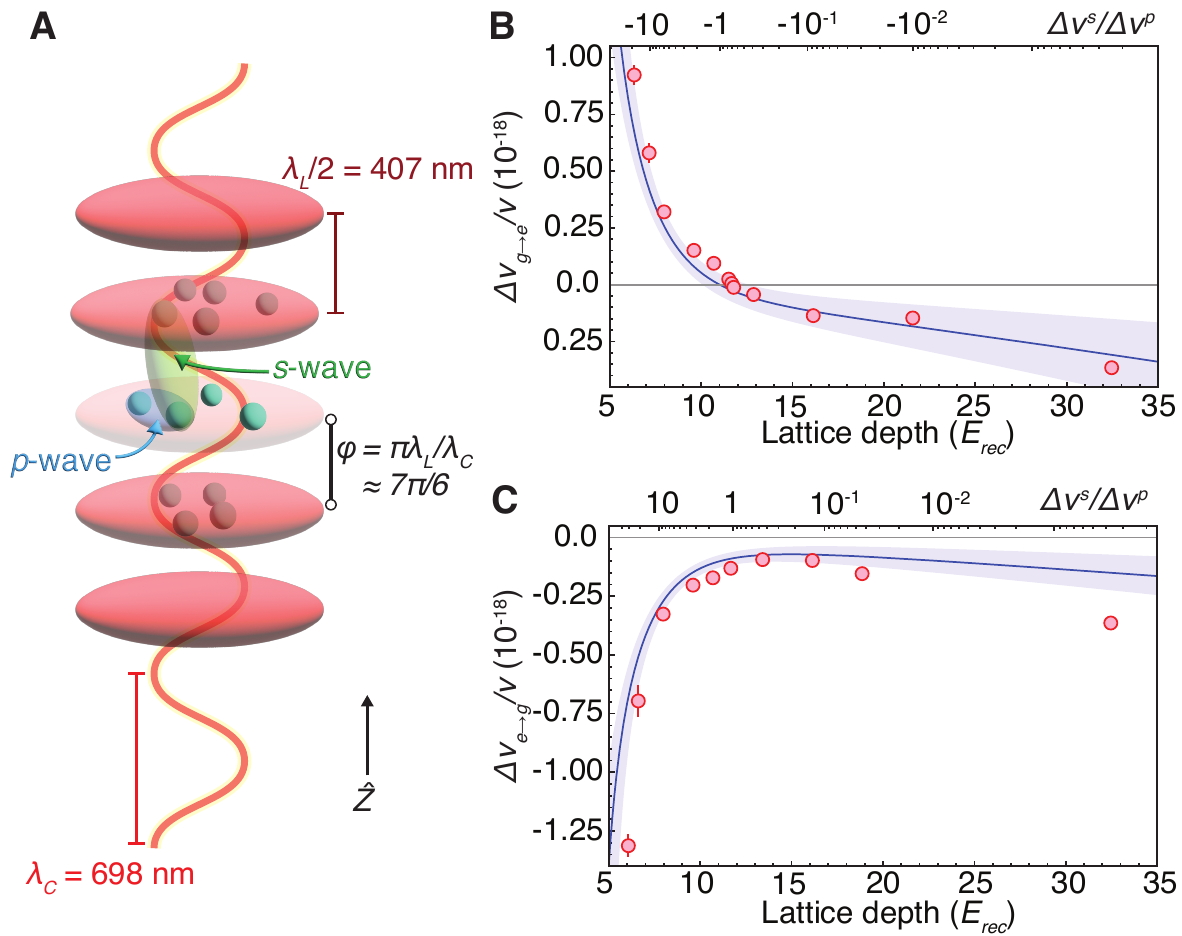}
\caption{\textbf{Engineering Interactions.} (\textbf{A}) By varying the lattice depth during clock spectroscopy, we modify the ratio of  off-site $s$-wave to on-site $p$-wave collisional shifts, $\Delta\nu^{s}_{\alpha\to\beta}/\Delta\nu^{p}_{\alpha\to\beta}$, where $\alpha$ and $\beta$ indicate clock states.
Atoms are trapped in an optical lattice with wavelength $\lambda_L$ and probed by clock light with wavelength $\lambda_C$. 
Each antinode of the lattice light traps a number of atoms which interact via $p$-wave collisions.
The $698$ nm clock wavelength is incommensurate with the lattice spacing, so atoms in neighboring lattice sites see different clock phases, $\varphi = \pi \lambda_L/\lambda_C \approx 7 \pi / 6$, allowing $s$-wave interactions at low lattice depths.
(\textbf{B}) The fractional frequency density shift $\Delta \nu_{g\to e}/\nu$ over a range of lattice depths. 
Red points and error bars indicate experimental data and corresponding uncertainty in density shift and lattice depth.
The theoretical density shift is shown as a solid blue line with the shaded blue region accounting for  uncertainties in the $s$-wave scattering length and $p$-wave scattering volumes~\cite{zhang2014spectroscopic,goban2018emergence}, as well as $10$ nK temperature uncertainty (details in SOM).
(\textbf{C}) The density shift  $\Delta \nu_{e\to g}/\nu$ over a range of lattice depths.  
}
\label{fig_2}
\end{figure}

\begin{figure}[htp!]
\centering
\includegraphics[width=5.5 cm]{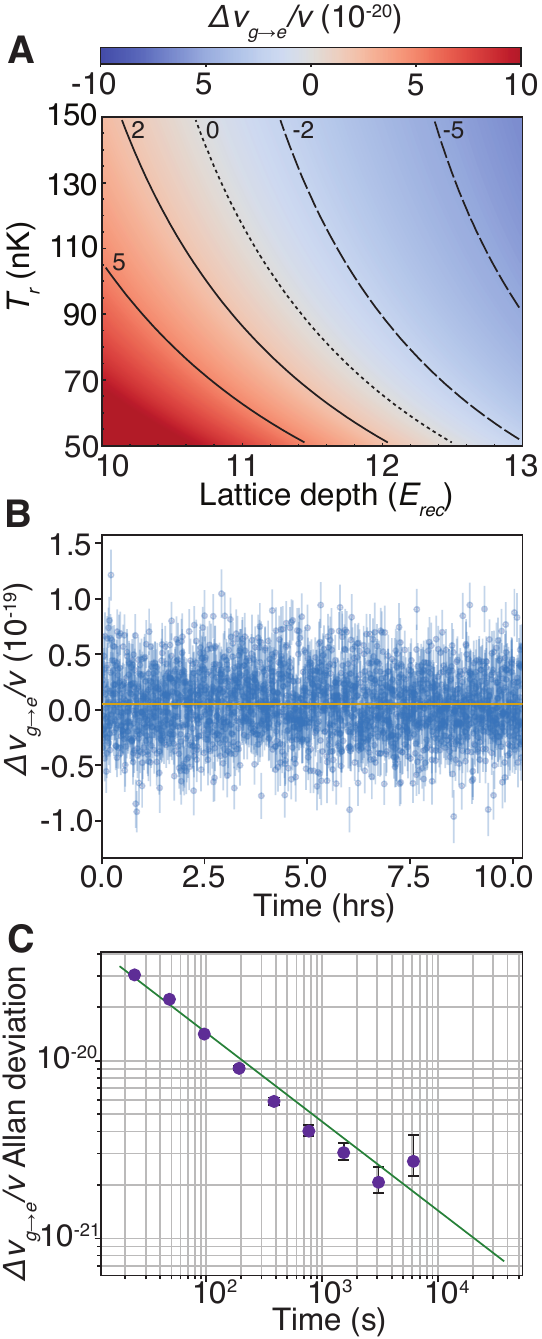}
\caption{\textbf{Density Shift Cancellation.} 
(\textbf{A}) At roughly $12$ $E_{rec}$, the contributions from $p$-wave and $s$-wave interactions balance, allowing clock operation with a density shift many orders lower than previous regimes.
The heat map and contours show the calculated the fractional frequency shift coefficient $\Delta \nu_{g\to e}/ \nu$ for a range of radial temperatures $T_r$ and lattice depths in our system. 
(\textbf{B})
Over a $10$ hour measurement, we report a mean coefficient $\Delta \nu_{g\to e}/\nu = 5.0(1.7) \times 10^{-21}$.
For each set of four lock points, we extract a density shift coefficient, shown with corresponding uncertainty in blue.
The weighted mean over the duration of the run is in gold.
(\textbf{C}) Allan deviation of the density shift coefficient (purple dots) with corresponding uncertainty reported in error bars. The green line is an instability fit with slope $1.3\times 10^{-19} / \sqrt{\tau}$ for averaging time $\tau$.}
\label{fig_3}
\end{figure}
\paragraph*{Density shifts in the carrier transition}
To measure the effect of collisional shifts on the clock transition, we perform extended measurements using a `clock lock' to track the drift of the laser. Each clock lock consists of a set of four lock points, a standard interleaved sequence probing opposite sign $m_F$ states to reject first order Zeeman shifts.  As reported in~\cite{bothwell2021}, we employ in situ imaging to construct a microscopic frequency map throughout the extended sample, fitting a linear slope to the relationship between frequency and number of atoms per site at each lock point. 
We define a linear density shift coefficient $\Delta \nu_{\alpha\to\beta}/\nu$ such that the total fractional frequency shift is the product of this coefficient and $N_{loc}$, calibrated using quantum projection noise techniques.
The reported values of $\Delta \nu_{\alpha\to\beta}/\nu$ are  the weighted mean of $\Delta \nu_{\alpha\to\beta}/\nu$ at every lock point during  an extended clock lock measurement campaign. The statistical uncertainty is given by the Allan deviation fit at $1/6$ total measuring time. 

In Fig.~\ref{fig_2} we plot the measured coefficients over a range of $V_0$ for both the $|g \rangle \rightarrow |e \rangle$ and $|e \rangle \rightarrow |g \rangle$ transition.
We typically utilize a $3.2$ s $\pi$ pulse duration.
To account for increased delocalization and reduced Rabi frequencies at the shallowest depths we utilize longer pulses. The effect of $s$-wave collisions at low lattice depths is readily apparent, with a dramatic increase in density shift over many orders in magnitude between $12$ $E_{rec}$ and $5$ $E_{rec}$, consistent with the growth of the off-site matrix element $\eta_1$ as $V_0$ is reduced. For the $|g \rangle \rightarrow |e \rangle$ transition presented in Fig.~\ref{fig_2}B, the $s$-wave frequency shift has an opposite sign compared to that of the $p$-wave. At the magic lattice depth, the $s$-wave and $p$-wave shifts have the same magnitude, resulting in a nearly perfect cancellation for a vanishingly small collisional frequency shift.
In the $|e \rangle \rightarrow |g \rangle$ case presented in Fig.~\ref{fig_2}C, the $s$-wave frequency shift has the same sign as that of the $p$-wave, and thus the density shift remains negative over all lattice depths.  
This behavior is well described by the mean-field solution from Eq.~(\ref{MFS}), represented by the solid blue lines in Fig.~\ref{fig_2}B and Fig.~\ref{fig_2}C. 
The disagreement at a large $V_0$ of $32$ $E_{rec}$, as shown in Fig.~\ref{fig_2}C, likely arises from increased light scattering not included in our theoretical model.

In Fig.~\ref{fig_3}A, we model the fractional frequency shift over a range of experimentally relevant lattice depths and radial temperatures near this magic point.
The density shift is sensitive to ensemble temperature, lattice depth, and excitation fraction.
Experimentally, the lattice depth is maintained through a precise and large bandwidth lattice intensity servo, and our clock lock tracks the laser drift to ensure a similar excitation fraction throughout the measurement duration.
The atomic temperature is less precisely controlled, with small drifts in the cooling laser frequency and stray magnetic fields contributing to reduced cooling reproducibility and observed $10$ nK variation. To evaluate the robustness of operating at the magic lattice depth, we demonstrate a $10$ hour clock lock using a $3.2$ s Rabi probe near the magic depth and report a $5.0(1.7) \times 10^{-21}$ fractional frequency shift per atom, as shown in Fig. \ref{fig_3}B. 
There is no apparent long term trend in the density shift, and the coefficient seems to reach a flicker beyond $\sim$1000 s, as shown by the Allan deviation in Fig. \ref{fig_3}C.

The data presented in Fig.~\ref{fig_3} was collected in a relatively high atom number  configuration.
For comparison, the synchronous measurement presented in  \cite{bothwell2021} with single clock instability of $3.1 \times 10^{-18}$ at $1$ s utilized $0.5$ mm length samples with an average of $38$ atoms per site. 
Operating in the density shift regime near the magic lattice depth presented here, the average density shift magnitude would be approximately $1.9(0.6) \times 10^{-19}$. 

\begin{figure}[tph!]
\centering
\includegraphics[width=16 cm]{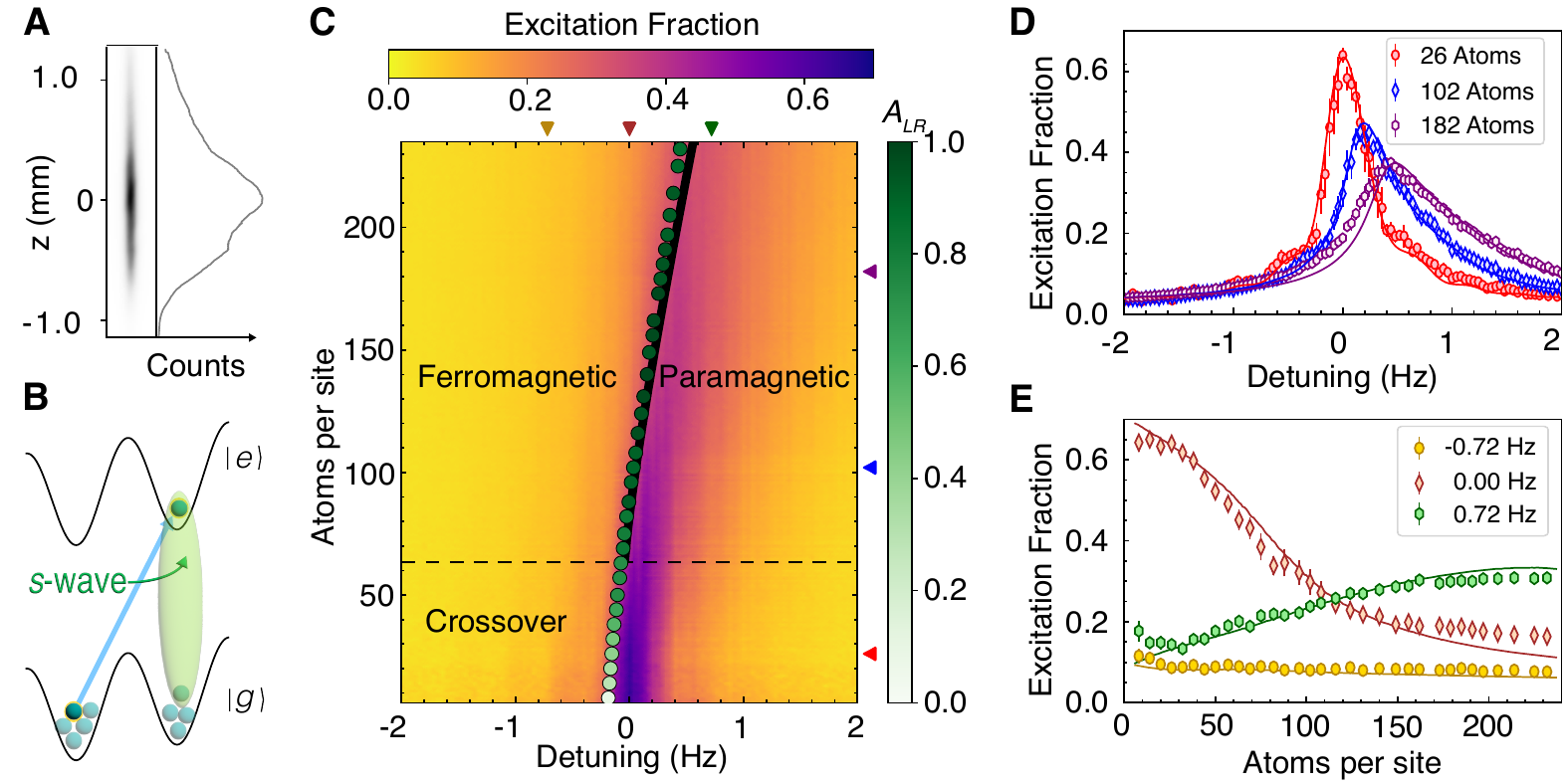}
\caption{\textbf{Dynamical Phase Transition.} (\textbf{A}) An image of lattice trapped atoms, indicating a spatial extent over a millimeter in length.
Within a single image we can study lattice site density regimes ranging over two orders of magnitude, shown here as camera counts.
(\textbf{B}) Addressing the $|g \, ; \, W_n\rangle \rightarrow |e \, ; \, W_{n + 1} \rangle$ transition, $s$-wave interactions effectively become on-site, leading to a strong collisional shift.
(\textbf{C}) The excitation fraction as a function of detuning and atom number on the $|g \, ; \, W_m\rangle \rightarrow |e \, ; \, W_{m + 1} \rangle$ transition at $22$ $E_{rec}$.
Above $\sim 63$ atoms per site, denoted by the dashed black line, the system features a dynamical phase transition between ferromagnetic and paramagnetic phases when varying the laser detuning and atomic density. The phase boundary is denoted by a solid black line from theoretical calculations and green points from the experimental data. The normalized asymmetry of the lineshape $A_{LR}$ is indicated by the shade of these points.
Arrows on the right and top axis indicate data plotted in \textbf{D} and \textbf{E} at constant atom number and detuning.
(\textbf{D}) Excitation fraction as a function of detuning at different atom numbers demonstrates the significant distortion and asymmetry that arises in the strongly interacting regime.
(\textbf{E}) Excitation fraction as a function of atom number in the ferromagnetic phase ($-0.72$ Hz), across the phase transition ($0$ Hz), and in the paramagnetic phase ($0.72$ Hz). 
}
\label{fig_4}
\end{figure}

\paragraph*{Dynamical Phase Transition} 
By addressing a transition to a different WS state, we further modify the atomic interactions. We can still define an interaction spin model by identifying the states $|\uparrow_{\bf{n}}\rangle\equiv|e;n_X,n_Y,W_{n+l}\rangle$ and $|\downarrow_{\bf{n}}\rangle\equiv|g;n_X,n_Y,W_{n}\rangle$ as the spin-1/2 internal levels. In particular, we interrogate the $l = 1$ transition.   The many-body dynamics are then described by the same mean field equation of motion, Eq. \ref{MFS}, but with a different effective magnetic field (see SOM):
\begin{equation}
    \mathbf{B}_{l=1}^{\perp}\approx\{\Omega_1,0,-\delta_{1}+2\chi_1^{\l=1}\langle S^z\rangle\},
    \label{eq:b2}
\end{equation} 
where $\chi_1^{l=1}=-\eta_0 U_{eg}/2$ and $\delta_{1}$ is the detuning of the laser to the $l=1$ transition. Note that because the wavefunction of the excited state is displaced by one lattice site (see Fig. \ref{fig_4}B), the overlap matrix element that characterizes the $s$-wave interactions is proportional to $\eta_0$.
Therefore atomic interactions are significantly enhanced in this case and increase with higher trap depth. Although the SOC phase does not enter directly in $\chi_1^{l=1} $, SOC still plays a key role by allowing the transition to be driven, see Eq. (\ref{omega}).  The stronger interactions modify the spin dynamics more dramatically and give rise to a dynamical phase transition (DPT) between dynamical ferromagnetic and paramagnetic phases (see SOM).  The DPT appears as a sharp change in behavior of the long-time average excitation fraction for an initial state prepared with all atoms in $|g \rangle$, $\overline{n_{\uparrow}}=\lim_{T\rightarrow\infty}\frac{1}{T}\int_0^T n_{\uparrow}(t)\mathrm{d}t$. 
In the dynamical ferromagnetic phase, interactions dominate and the system features small oscillations near a single pole of the Bloch sphere, with   $\overline{n_{\uparrow}}\approx 0$. In the  dynamical paramagnetic phase,  the system exhibits large excursions around the Bloch sphere and  $\overline{n_{\uparrow}}$ dynamically adjusts itself as ${\delta_{1}}$ is varied. 
In the interaction dominant regime, the DPT generates a second order critical line that distinguishes the two dynamical phases. The transition evolves into a smooth crossover region in the weakly interacting regime, where the dynamics are dominated by single-particle Rabi flopping.

Similar to other DPT experiments, instead  of direct measurements of $\overline{n_{\uparrow}}$, the order parameter is estimated by measuring the excitation fraction at a fixed probe time. We use a $2.3$ s Rabi $\pi$ pulse with lattice depth $V_0=22$ $E_{rec}$ and radial temperature $T_r=190$ nK. Within a single image we observe a density range spanning over two orders of magnitude (Fig. \ref{fig_4}A).  We spatially resolve the excitation fraction   within the sample and construct the dynamical phase diagram shown in Fig.~\ref{fig_4}C.

For a given $N_{\rm loc}$ we extract the lineshape asymmetry $A_{LR}$ defined as $(n_R-n_L)/(n_R+n_L)$ from experimental data, and normalize by the maximum value of $A_{LR}$. Here, $n_R=\int_{\delta_{max}}^{\delta_{max}+f} n_{\uparrow}(\delta)\mathrm{d}\delta$, $n_L=\int_{\delta_{max}-f}^{\delta_{max}} n_{\uparrow}(\delta)\mathrm{d}\delta$, where $\delta_{max}$ is the detuning for the peak value of the Rabi lineshape, and $f/2\pi=1$ Hz covers almost the entire frequency range of the Rabi lineshape. The lineshape asymmetry allows us to characterize the dynamical phases. 
For $N_{loc}<63$, bellow the dashed black line in Fig. \ref{fig_4}C, the system is in a crossover regime  featuring a linear density shift and asymmetry $A_{LR}$ that becomes more pronounced as the atom number increases.
Over $63$ atoms per site, the lineshape is near maximally asymmetric, and distinct ferromagnetic and paramagnetic dynamical phases are identified.   
The phase boundary is experimentally determined by finding the maximum derivative of the lineshape as a function of detuning, plotted as green points in Fig.~\ref{fig_4}C, with $A_{LR}$ indicated by the shade. The points lie very close to the theoretically calculated  phase  boundary  shown as a solid black line. 

The asymmetry in the lineshape becomes apparent by viewing the excitation at a constant atom number, as in Fig.~\ref{fig_4}D. At densities well below the crossover boundary, the lineshape is only slightly distorted from that of an ideal Rabi response. Above the crossover density, the excitation displays very different behaviors for the two opposite signs of detuning, and the excitation becomes highly insensitive to changes of detuning deep in the ferromagnetic phase. The constant detuning profiles presented in Fig.~\ref{fig_4}E further illustrate this dynamical phase transition. At $\delta_{1}/2\pi=0$ Hz, the laser drive is on resonance with the non-interacting transition. Above  the crossover regime, the ensemble features  both dynamical phases, evolving from a dynamical paramagnet to a dynamical ferromagnet for  $N_{loc}  >82$. 
At $\delta_{1}/2\pi=-0.72$ Hz the system is in the dynamical ferromagnetic phase  above the crossover region. 
However, with $\delta_{1}/2\pi=0.72$ Hz detuning, the excitation fraction initially rises with atom number when the system is in the paramagnetic phase and saturates close to the phase boundary. In both panels D and E the solid lines are theoretical predictions from the mean field spin model with an additional dephasing term accounting for mode-changing collisions (see SOM).

\paragraph*{Conclusions and Outlook}
Operating in the Wannier-Stark regime has realized a new and optimized platform for optical lattice clocks, with record coherence time and clock precision~\cite{bothwell2021}. The work here highlights the use of Hamiltonian engineering and control of atomic interactions to remove the compromise between increased precision and reduced systematic uncertainties. Operating with hundreds of thousands of atoms we still limit the density-related frequency shift well below the current state of the art, and further reduction in density shift is readily attainable. Importantly, this work utilizes precise tuning of interactions to explore rich many-body behavior. With selective  Wannier-Stark interrogation and in situ imaging, we efficiently map out a dynamical phase transition over a range of density of more than two orders of magnitude. 

So far we operate in a regime where a mean-field model is sufficient to describe the many-body dynamics. Driving the system with more sophisticated pulse sequences will allow us to further explore quantum correlation and beyond mean-field effects. This will  open a path for the generation of spin squeezed states with a net quantum metrological advantage for state-of-the-art quantum sensors.

\bibliography{main}

\bibliographystyle{Science}

\paragraph*{Acknowledgments}
We thank K. Kim, J. Zaris, R. Hutson, J. Robinson, C. Sanner, A. Staron, W. Milner, J. Meyer, and E. Oelker for stimulating discussions and experimental contributions. We thank K. Kim and W. McGrew for careful reading of the manuscript. Funding for this work was provided by DARPA, ARO (W911NF-16-1-0576), AFOSR (FA9550-18-1-0319, FA9550-19-1-027), NSF QLCI OMA–2016244, NSF Phys-1734006, DOE National Quantum Information Science Research Centers (Quantum Systems Accelerator), and NIST. 

\paragraph*{Author Contributions} The experiment was performed by A.A., T.B., C.J.K., D.K., and J.Y. The theory model was established by A.C., P.H. and A.M.R. All authors contributed to analyzing the results and writing the manuscript.

\paragraph*{Competing Interests}
The authors declare no competing interests.

\paragraph*{Data and materials availability}
 All data in the manuscript and supplementary materials is available upon request.

\newpage

\section*{Supplementary materials}

\setcounter{equation}{0}
\setcounter{figure}{0}
\setcounter{table}{0}
\makeatletter
\renewcommand{\theequation}{S\arabic{equation}}
\renewcommand{\thefigure}{S\arabic{figure}}
\renewcommand{\thesection}{S\arabic{section}}

\tableofcontents
\newpage

\section{Theoretical Model}
\subsection{SU($N$) interactions in the Sr optical lattice clock}
Fermionic $^{87}$Sr atoms  have  two long-lived electronic orbitals, the ${}^1S_0$ and ${}^3P_0$ clock states, as well as a nuclear spin degree of freedom  with $I=9/2$. We denote the electronic states as $|g\rangle$ and $|e\rangle$ respectively and the  $N=2I+1$ nuclear spin levels  as  $m=-I,-I+1,\cdots, I$. Due to the lack of hyperfine coupling between the nuclear and electronic degrees of freedom, the scattering parameters that describe   two-body interactions are independent of the nuclear spin states. This  property gives rise to an interaction Hamiltonian invariant under SU($N$) rotations \cite{gorshkov2010twoorbital,zhang2014spectroscopic}.

$S$-wave interactions occur under spatially symmetric collisions. Due to  the requirement  for fermionic atoms  to feature   a fully  antisymmetric total wave function, to collide under the  $s$-wave  channel  symmetric nuclear spin states require their electronic orbitals to be antisymmetric. Therefore,  the state $(|ge\rangle-|eg\rangle)/\sqrt{2}$  is the only one that can feature $s$-wave interactions,  characterized  by  the $s$-wave scattering length $a_{eg}^{-}$. Similarly, to collide via $s$-wave interactions, antisymmetric nuclear spin states require their electronic orbitals to be symmetric. So there are three possible combination of  electronic states  $|gg\rangle$, $|ee\rangle$, $(|ge\rangle+|eg\rangle)/\sqrt{2}$ that can collide via $s$-wave.  Their interactions  are characterized by  the $s$-wave scattering lengths $a_{gg}$, $a_{ee}$ and $a_{eg}^{+}$ respectively.

Defining the permutationally symmetric ( $\mathcal{P}_{+}$) and  permutationally antisymmetric ( $\mathcal{P}_{-}$) projector operators,  $\mathcal{P}_{\pm}=(\mathcal{I}\pm\mathcal{P}_{12})/2$ with $\mathcal{P}_{12}=\sum_{mm'}|m\rangle_1\langle m'| \otimes |m'\rangle_2\langle m|$, where $m,m'$ label nuclear spin levels, the resulting  $s$-wave pseudopotential for SU($N$) interaction takes the following form: $ V_s^{gg}(\mathbf{R}_{12})\propto a_{gg}\mathcal{P}_{-}$, $V_s^{ee}(\mathbf{R}_{12})\propto a_{ee}\mathcal{P}_{-}$, $V_s^{eg}(\mathbf{R}_{12})\propto (a_{eg}^{-}+a_{eg}^{+})\mathcal{I}/2+(a_{eg}^{-}-a_{eg}^{+})\mathcal{P}_{12}/2$.
Thus the $s$-wave interaction Hamiltonian in the second quantized form,
\begin{equation}
    \begin{aligned}
    H_{s}&=\frac{2\pi\hbar^2a_{gg}}{M}\sum_{\substack{mm'\\(m\neq m')}}\int\mathrm{d}^3\mathbf{R}\,\psi^{\dag}_{gm}(\mathbf{R})\psi^{\dag}_{gm'}(\mathbf{R})\psi_{gm'}(\mathbf{R})\psi_{gm}(\mathbf{R})\\
    &+\frac{2\pi\hbar^2a_{ee}}{M}\sum_{\substack{mm'\\(m\neq m')}}\int\mathrm{d}^3\mathbf{R}\,\psi^{\dag}_{em}(\mathbf{R})\psi^{\dag}_{em'}(\mathbf{R})\psi_{em'}(\mathbf{R})\psi_{em}(\mathbf{R})\\
    &+\frac{2\pi\hbar^2(a_{eg}^{-}+a_{eg}^{+})}{M}\sum_{mm'}\int\mathrm{d}^3\mathbf{R}\,\psi^{\dag}_{gm}(\mathbf{R})\psi^{\dag}_{em'}(\mathbf{R})\psi_{em'}(\mathbf{R})\psi_{gm}(\mathbf{R})\\
    &+\frac{2\pi\hbar^2(a_{eg}^{-}-a_{eg}^{+})}{M}\sum_{mm'}\int\mathrm{d}^3\mathbf{R}\,\psi^{\dag}_{gm}(\mathbf{R})\psi^{\dag}_{em'}(\mathbf{R})\psi_{em}(\mathbf{R})\psi_{gm'}(\mathbf{R}),\\
    \end{aligned}
    \label{eq:swave}
\end{equation}
where $M$ is the mass of a Sr atom, $\psi_{gm}(\mathbf{R})$ and $\psi_{em'}(\mathbf{R})$ are fermionic annihilation field operators of nuclear spin $m$ in ground manifold and nuclear spin $m'$ in excited manifold respectively.  

For spatially antisymmetric $p$-wave interactions, antisymmetric nuclear spin states require their electronic orbitals to be antisymmetric for an antisymmetric total wavefunction, so the only possible electronic state is $(|ge\rangle-|eg\rangle)/\sqrt{2}$, which interacts via  the $p$-wave scattering volume $(b_{eg}^{-})^3$. Symmetric nuclear spin states require their electronic orbitals to be symmetric, so the three possible electronic states are $|gg\rangle$, $|ee\rangle$, $(|ge\rangle+|eg\rangle)/\sqrt{2}$, which interact via the $p$-wave scattering volumes $b_{gg}^3$, $b_{ee}^3$ and $(b_{eg}^{+})^3$ respectively. The $p$-wave pseudopotential for SU($N$) interaction takes the following form: $V_p^{gg}(\mathbf{R}_{12})\propto b_{gg}^3\mathcal{P}_{+}$, $V_p^{ee}(\mathbf{R}_{12})\propto b_{ee}^3\mathcal{P}_{+}$,
$V_p^{eg}(\mathbf{R}_{12})\propto [(b_{eg}^{+})^3+(b_{eg}^{-})^3]\mathcal{I}/2+[(b_{eg}^{+})^3-(b_{eg}^{-})^3]\mathcal{P}_{12}/2$.
This leads to the $p$-wave interaction Hamiltonian in the second quantized form,
\begin{equation}
    \begin{aligned}
    H_p&=\frac{3\pi\hbar^2b_{gg}^3}{2M}\sum_{mm'}\int\mathrm{d}^3\mathbf{R}\,[(\nabla\psi^{\dag}_{gm})\psi^{\dag}_{gm'}-\psi^{\dag}_{gm}(\nabla\psi^{\dag}_{gm'})]\cdot [\psi_{gm'}(\nabla\psi_{gm})-(\nabla\psi_{gm'})\psi_{gm}]\\
    &+\frac{3\pi\hbar^2b_{ee}^3}{2M}\sum_{mm'}\int\mathrm{d}^3\mathbf{R}\,[(\nabla\psi^{\dag}_{em})\psi^{\dag}_{em'}-\psi^{\dag}_{em}(\nabla\psi^{\dag}_{em'})]\cdot [\psi_{em'}(\nabla\psi_{em})-(\nabla\psi_{em'})\psi_{em}]\\
    &+\frac{3\pi\hbar^2[(b_{eg}^{+})^3+(b_{eg}^{-})^3]}{2M}\sum_{mm'}\int\mathrm{d}^3\mathbf{R}\,[(\nabla\psi^{\dag}_{gm})\psi^{\dag}_{em'}-\psi^{\dag}_{gm}(\nabla\psi^{\dag}_{em'})]\cdot [\psi_{em'}(\nabla\psi_{gm})-(\nabla\psi_{em'})\psi_{gm}]\\
    &+\frac{3\pi\hbar^2[(b_{eg}^{+})^3-(b_{eg}^{-})^3]}{2M}\sum_{mm'}\int\mathrm{d}^3\mathbf{R}\,[(\nabla\psi^{\dag}_{gm})\psi^{\dag}_{em'}-\psi^{\dag}_{gm}(\nabla\psi^{\dag}_{em'})]\cdot [\psi_{em}(\nabla\psi_{gm'})-(\nabla\psi_{em})\psi_{gm'}].
    \end{aligned}
    \label{eq:pwave}
\end{equation}

We focus on the least magnetically sensitive clock transition in $^{87}$Sr, $|{}^{1}S_0,m_F=\pm 5/2\rangle\rightarrow |{}^{3}P_0,m_F=\pm 3/2\rangle$, denoted by $|\tilde{g}\rangle$ and $|\tilde{e}\rangle$ respectively. In a large magnetic field, the flip-flop process of nuclear spin states in Eq.~(\ref{eq:swave}) and Eq.~(\ref{eq:pwave}) can be ignored, so the interaction Hamiltonian including $s$-wave and $p$-wave contributions can be restricted to these two states, so
\begin{equation}
    \begin{aligned}
    H_{\mathrm{int}}&=\frac{2\pi\hbar^2(a_{eg}^{-}+a_{eg}^{+})}{M}\int\mathrm{d}^3\mathbf{R}\,\psi^{\dag}_{\tilde{e}}(\mathbf{R})\psi^{\dag}_{\tilde{g}}(\mathbf{R})\psi_{\tilde{g}}(\mathbf{R})\psi_{\tilde{e}}(\mathbf{R})\\ 
    &+\frac{3\pi\hbar^2b_{gg}^3}{2M}\int\mathrm{d}^3\mathbf{R}\,[(\nabla\psi^{\dag}_{\tilde{g}})\psi^{\dag}_{\tilde{g}}-\psi^{\dag}_{\tilde{g}}(\nabla\psi^{\dag}_{\tilde{g}})]\cdot [\psi_{\tilde{g}}(\nabla\psi_{\tilde{g}})-(\nabla\psi_{\tilde{g}})\psi_{\tilde{g}}]\\
    &+\frac{3\pi\hbar^2b_{ee}^3}{2M}\int\mathrm{d}^3\mathbf{R}\,[(\nabla\psi^{\dag}_{\tilde{e}})\psi^{\dag}_{\tilde{e}}-\psi^{\dag}_{\tilde{e}}(\nabla\psi^{\dag}_{\tilde{e}})]\cdot [\psi_{\tilde{e}}(\nabla\psi_{\tilde{e}})-(\nabla\psi_{\tilde{e}})\psi_{\tilde{e}}]\\
    &+\frac{3\pi\hbar^2[(b_{eg}^{+})^3+(b_{eg}^{-})^3]}{2M}\int\mathrm{d}^3\mathbf{R}\,[(\nabla\psi^{\dag}_{\tilde{g}})\psi^{\dag}_{\tilde{e}}-\psi^{\dag}_{\tilde{g}}(\nabla\psi^{\dag}_{\tilde{e}})]\cdot [\psi_{\tilde{e}}(\nabla\psi_{\tilde{g}})-(\nabla\psi_{\tilde{e}})\psi_{\tilde{g}}].
    \end{aligned}
    \label{eq:int}
\end{equation}

\subsection{Spin model for the carrier transition}
As described in the main text, our experimental system is a vertical 1D lattice with magic wavelength ($\lambda_L=813$~nm), so the external trapping potential $V_{\mathrm{ext}}(\mathbf{R})$ is the same for $|\tilde{g}\rangle$ and $|\tilde{e}\rangle$ states. To the leading order, we have
\begin{equation}
    V_{\mathrm{ext}}(\mathbf{R})\approx V_0\sin^2(k_LZ)+MgZ+\frac{1}{2}M\omega_R^2(X^2+Y^2).
\end{equation}
Here, $k_L=2\pi/\lambda_L$ is the wavenumber of the lattice that sets the atomic recoil energy $E_{rec}=\hbar^2k_L^2/2M$, $g$ is the gravitational acceleration, and $\omega_R$ is the radial trapping frequency. In addition, the clock laser ($\lambda_c=698$~nm), aligned with the lattice direction, drives the transitions between $|\tilde{g}\rangle$ and $|\tilde{e}\rangle$ states with bare Rabi frequency $\Omega$ and detuning $\delta$. In the rotating frame of the clock laser, the second quantized Hamiltonian take the following form,
\begin{equation}
    H=H_0+H_{\mathrm{int}}+H_{\mathrm{laser}},
\end{equation}
where
\begin{equation}
    H_0=\sum_{\alpha=\{\tilde{g},\tilde{e}\}}\int\mathrm{d}^3\mathbf{R}\,\psi^{\dag}_{\alpha}(\mathbf{R})\bigg[-\frac{\hbar^2}{2M}\nabla^2+V_{\mathrm{ext}}(\mathbf{R})\bigg]\psi_{\alpha}(\mathbf{R}),
\end{equation}
$H_{\mathrm{int}}$ is given by Eq.~(\ref{eq:int}), and 
\begin{equation}
    H_{\mathrm{laser}}=\frac{\hbar\Omega}{2}\int\mathrm{d}^3\mathbf{R}\bigg[\mathrm{e}^{ik_cZ}\psi^{\dag}_{\tilde{e}}(\mathbf{R})\psi_{\tilde{g}}(\mathbf{R})+\mathrm{h.c.}\bigg]-\frac{\hbar\delta}{2}\int\mathrm{d}^3\mathbf{R}\bigg[\psi^{\dag}_{\tilde{e}}(\mathbf{R})\psi_{\tilde{e}}(\mathbf{R})-\psi^{\dag}_{\tilde{g}}(\mathbf{R})\psi_{\tilde{g}}(\mathbf{R})\bigg].
\end{equation}
Here, $k_c=2\pi/\lambda_c$ is the wavenumber of the clock laser, and $\psi_{\alpha}(\mathbf{R})$ is the annihilation field operator for a fermionic atom of internal state $\alpha$. 

Our experiment operates in the regime where the collisional rate of relaxation for motional degrees of freedom  is slower than internal spin dynamics and trapping frequencies \cite{martin2013,rey2014,zhang2014spectroscopic,smale2019,Chu2020}.
This  condition ensures  only internal levels evolve while atoms  remain frozen   in their single-particle eigenstates during the dynamics. We first focus on the case of the  carrier transition, where the clock laser couples the following two single particle states:  $\mathbf{n}$, $|\uparrow_{\mathbf{n}}\rangle\equiv|\tilde{e};n_X,n_Y,W_n\rangle$ and $|\downarrow_{\mathbf{n}}\rangle\equiv|\tilde{g};n_X,n_Y,W_n\rangle$, where $\mathbf{n}=\{n_X,n_Y,n\}$, with $n_X, n_Y$ denoting  the radial harmonic oscillator modes and $n$ the lattice site index of the center of the Wannier-Stark state $|W_n\rangle$.
We expand the field operator $\psi_{\alpha}(\mathbf{R})$ in terms of single-particle eigenstates as follows,
\begin{equation}
    \psi_{\tilde{e}}(\mathbf{R})=\sum_{\mathbf{n}}\phi_{n_X}(X)\phi_{n_Y}(Y)W_n(Z)c_{\mathbf{n}\uparrow}, \quad  \psi_{\tilde{g}}(\mathbf{R})=\sum_{\mathbf{n}}\phi_{n_X}(X)\phi_{n_Y}(Y)W_n(Z)c_{\mathbf{n}\downarrow},
\end{equation}
where $c_{\mathbf{n}\uparrow}$ and $c_{\mathbf{n}\downarrow}$ are fermionic annihilation operators for $|\uparrow_{\mathbf{n}}\rangle$ and $|\downarrow_{\mathbf{n}}\rangle$ states respectively. Here, the harmonic oscillator wave function is
\begin{equation}
    \phi_{n_{X}}(X)=\frac{1}{\sqrt{2^{n_{X}} n_{X}!}}\bigg(\frac{M\omega_R}{\pi\hbar}\bigg)^{1/4}\mathrm{e}^{-M\omega_R X^2/2\hbar}H_{n_{X}}\bigg(\sqrt{\frac{M\omega_r}{\hbar}}X\bigg),
\end{equation}
where $H_{n_{X}}(X)$ are Hermite polynomials. The wave function for the Wannier-Stark state is
\begin{equation}
    W_n(Z)=\sum_{m}\mathcal{J}_{m-n}\bigg(\frac{2J_0}{Mga_l}\bigg)w(Z-ma_L),
    \label{eq:wannierstark}
\end{equation}
where $\mathcal{J}_{n}(x)$ are Bessel functions, $J_0\approx (4/\sqrt{\pi})E_{rec}^{1/4}V_0^{3/4}\exp[-2\sqrt{V_0/E_{rec}}]$ is the ground band nearest-neighbor tunneling energy, $a_L=\lambda_L/2$ is the lattice spacing, and $w(Z)$ is the ground band Wannier function centering at $Z=0$.

Under the frozen mode approximation, we treat each atom as a spin-$1/2$ system spanned by $|\uparrow_{\mathbf{n}}\rangle$ and $|\downarrow_{\mathbf{n}}\rangle$ states. Therefore, we define the spin operators,
\begin{equation}
    \begin{gathered}
        S^x_{\mathbf{n}}=\frac{1}{2}(c^{\dag}_{\mathbf{n}\uparrow}c_{\mathbf{n}\downarrow}+c^{\dag}_{\mathbf{n}\downarrow}c_{\mathbf{n}\uparrow}), \quad S^y_{\mathbf{n}}=-\frac{i}{2}(c^{\dag}_{\mathbf{n}\uparrow}c_{\mathbf{n}\downarrow}-c^{\dag}_{\mathbf{n}\downarrow}c_{\mathbf{n}\uparrow}),\\
        S^z_{\mathbf{n}}=\frac{1}{2}(c^{\dag}_{\mathbf{n}\uparrow}c_{\mathbf{n}\uparrow}-c^{\dag}_{\mathbf{n}\downarrow}c_{\mathbf{n}\downarrow}), \quad N_{\mathbf{n}}=c^{\dag}_{\mathbf{n}\uparrow}c_{\mathbf{n}\uparrow}+c^{\dag}_{\mathbf{n}\downarrow}c_{\mathbf{n}\downarrow},\\
    \end{gathered}
\end{equation}
and  rewrite the interaction Hamiltonian, 
\begin{equation}
    H_{\mathrm{int}}/\hbar=\sum_{\substack{\mathbf{n}\mathbf{m}\\(\mathbf{n}\neq\mathbf{m})}}\bigg[J^{\perp}_{\mathbf{n}\mathbf{m}}\mathbf{S}_{\mathbf{n}}\cdot\mathbf{S}_{\mathbf{m}}+\chi_{\mathbf{n}\mathbf{m}}S^z_{\mathbf{n}}S^z_{\mathbf{m}}+\frac{C_{\mathbf{n}\mathbf{m}}}{2}(S^z_{\mathbf{n}}N_{\mathbf{m}}+N_{\mathbf{n}}S^z_{\mathbf{m}})\bigg],
\end{equation}
where
\begin{equation}
    \begin{gathered}
    J^{\perp}_{\mathbf{n}\mathbf{m}}=\eta_{|n-m|}(V^{eg}_{\mathbf{n}\mathbf{m}}-U^{eg}_{\mathbf{n}\mathbf{m}})/2,\quad \chi_{\mathbf{n}\mathbf{m}}=\eta_{|n-m|}(V^{ee}_{\mathbf{n}\mathbf{m}}+V^{gg}_{\mathbf{n}\mathbf{m}}-2V^{eg}_{\mathbf{n}\mathbf{m}})/2,\\ C_{\mathbf{n}\mathbf{m}}=\eta_{|n-m|}(V^{ee}_{\mathbf{n}\mathbf{m}}-V^{gg}_{\mathbf{n}\mathbf{m}})/2.
    \end{gathered}
\end{equation}
Here, $\eta_{|n-m|}$ is a dimensionless overlap integral of Wannier-Stark states defined as
\begin{equation}
    \eta_{|n-m|}=\frac{\lambda_L}{\sqrt{2\pi}}\bigg(\frac{V_0}{E_{rec}}\bigg)^{-1/4}\int\mathrm{d}Z\,[W_n(Z)]^2[W_m(Z)]^2.
    \label{eq:integral}
\end{equation}
$U^{\alpha\beta}_{\mathbf{n}\mathbf{m}}$ and $V^{\alpha\beta}_{\mathbf{n}\mathbf{m}}$ are $s$-wave and $p$-wave interaction parameters respectively ($\alpha,\beta=\{g,e\}$),
\begin{equation}
    \begin{gathered}
    U^{\alpha\beta}_{\mathbf{n}\mathbf{m}}=\frac{8\pi\hbar a_{\alpha\beta}}{M}s_{n_{x}m_{x}}s_{n_{y}m_{y}}\frac{k_L}{\sqrt{2\pi}}\bigg(\frac{V_0}{E_{rec}}\bigg)^{1/4},\\ V^{\alpha\beta}_{\mathbf{n}\mathbf{m}}=\frac{6\pi\hbar b_{\alpha\beta}^3}{M}(p_{n_{x}m_{x}}s_{n_{y}m_{y}}+s_{n_{x}m_{x}}p_{n_{y}m_{y}})\frac{k_L}{\sqrt{2\pi}}\bigg(\frac{V_0}{E_{rec}}\bigg)^{1/4},
    \end{gathered}
    \label{eq:spd}
\end{equation}
where $a_{eg}\equiv (a_{eg}^{+}+a_{eg}^{-})/2$, $b_{eg}^3\equiv [(b_{eg}^{+})^3+(b_{eg}^{-})^3]/2$, $s_{nm}=\int\mathrm{d}X\,[\phi_{n}(X)]^2[\phi_{m}(X)]^2$, and $p_{nm}=\int\mathrm{d}X\,[(\partial_X\phi_{n}(X))\phi_{m}(X)-\phi_{n}(X)(\partial_X\phi_{m}(X))]^2$. Note that in $V^{\alpha\beta}_{\mathbf{n}\mathbf{m}}$ we ignore the $p$-wave contributions in the $\hat{Z}$ direction, because its leading order terms are overlap matrix elements of gradients of Wannier functions in nearest-neighbor lattice sites based on the expansion in Eq.~(\ref{eq:wannierstark}). These matrix elements  are small for parameters used in the experiment. 

On the carrier transition, $H_{\mathrm{laser}}$ becomes
\begin{equation}
    H_{\mathrm{laser}}/\hbar= \frac{1}{2}\sum_{\mathbf{n}}(\Omega_{\mathbf{n}}S^{+}_{\mathbf{n}}+\mathrm{h.c.})-\delta \sum_{\mathbf{n}}S^z_{\mathbf{n}},
\end{equation}
where
\begin{equation}
    \Omega_{\mathbf{n}}=\Omega\int\mathrm{d}Z\,\mathrm{e}^{ik_cZ}[W_n(Z)]^2=\Omega_0\mathrm{e}^{in\varphi}.
\end{equation}
Here, $\varphi=k_ca_L=\pi\lambda_L/\lambda_c$ is the clock laser phase difference between nearest-neighbor Wannier-Stark states, generating spin-orbit coupling. The Rabi frequency for carrier transition is
\begin{equation}
    \Omega_0=\Omega\cdot \mathcal{C}\mathcal{J}_0\bigg(\frac{4J_0}{Mga_L}\sin(\varphi/2)\bigg),
\end{equation}
where $\mathcal{C}=\int\mathrm{d}Z\,\mathrm{e}^{ik_cZ}[w(Z)]^2\approx \exp\big[-\lambda_L^2/4\lambda_c^2\sqrt{V_0/E_{rec}}\big]$. The dependence of $\Omega_0$ on lattice depth $V_0$ is shown in Fig.~\ref{supp_1}(A). In the following discussions, it is convenient to remove the phase dependence on lattice sites in the $H_{\mathrm{laser}}$ term by a gauge transformation $\tilde{c}_{\mathbf{n}\uparrow}=\mathrm{e}^{-in\varphi}c_{\mathbf{n}\uparrow}$, $\tilde{c}_{\mathbf{n}\downarrow}=c_{\mathbf{n}\downarrow}$. Under the gauge transformation the spin operators become:
\begin{equation}
    \begin{gathered}
    \tilde{S}^x_{\mathbf{n}}=\cos(n\varphi)S^x_{\mathbf{n}}-\sin(n\varphi)S^y_{\mathbf{n}}, \quad\tilde{S}^y_{\mathbf{n}}=\sin(n\varphi)S^x_{\mathbf{n}}+\cos(n\varphi)S^y_{\mathbf{n}},\\
    \tilde{S}^z_{\mathbf{n}}=S^z_{\mathbf{n}}, \quad\tilde{N}_{\mathbf{n}}=N_{\mathbf{n}}.
    \end{gathered}
    \label{eq:gauge}
\end{equation}

Combining the discussions above, the effective Hamiltonian in the gauged frame becomes
\begin{equation}
    \begin{aligned}
    H/\hbar&=\sum_{\substack{\mathbf{n}\mathbf{m}\\(\mathbf{n}\neq\mathbf{m})}}\bigg[\tilde{J}^{\perp}_{\mathbf{n}\mathbf{m}}\tilde{\mathbf{S}}_{\mathbf{n}}\cdot\tilde{\mathbf{S}}_{\mathbf{m}}+\tilde{\chi}_{\mathbf{n}\mathbf{m}}\tilde{S}^z_{\mathbf{n}}\tilde{S}^z_{\mathbf{m}}+D_{\mathbf{n}\mathbf{m}}(\tilde{S}^x_{\mathbf{n}}\tilde{S}^{y}_{\mathbf{m}}-\tilde{S}^y_{\mathbf{n}}\tilde{S}^{x}_{\mathbf{m}})+\frac{C_{\mathbf{n}\mathbf{m}}}{2}(\tilde{S}^z_{\mathbf{n}}\tilde{N}_{\mathbf{m}}+\tilde{N}_{\mathbf{n}}\tilde{S}^z_{\mathbf{m}})\bigg]\\
    &-\hbar\delta\sum_{\mathbf{n}}\tilde{S}^z_{\mathbf{n}}+\hbar\Omega_0\sum_{\mathbf{n}}\tilde{S}^x_{\mathbf{n}},
    \end{aligned}
    \label{eq:hamil}
\end{equation}
where $\tilde{J}^{\perp}_{\mathbf{n}\mathbf{m}}=\cos[(n-m)\varphi]J^{\perp}_{\mathbf{n}\mathbf{m}}$, $\tilde{\chi}_{\mathbf{n}\mathbf{m}}=\chi_{\mathbf{n}\mathbf{m}}+J^{\perp}_{\mathbf{n}\mathbf{m}}-\tilde{J}^{\perp}_{\mathbf{n}\mathbf{m}}$, and $D_{\mathbf{n}\mathbf{m}}=-\sin[(n-m)\varphi]J^{\perp}_{\mathbf{n}\mathbf{m}}$. 

Now we discuss the dependence of interaction parameters $\tilde{J}^{\perp}_{\mathbf{n}\mathbf{m}}$, $\tilde{\chi}_{\mathbf{n}\mathbf{m}}$ and $D_{\mathbf{n}\mathbf{m}}$ on radial harmonic oscillator modes ($n_X,n_Y,m_X,m_Y$) and the distance along lattice direction ($|n-m|$). As reported in \cite{martin2013,zhang2014spectroscopic,Chu2020}, the overlap integrals  are not overly sensitive to the    radial modes in consideration, allowing  us to simplify  the Hamiltonian dynamics in terms of collective spin operators at each lattice site,
$S^{x,y,z}_n=\sum_{n_Xn_Y}\tilde{S}_{\mathbf{n}}^{x,y,z}$, $N_n=\sum_{n_Xn_Y}\tilde{N}_{\mathbf{n}}$. 
Due to the partial delocalization of the Wannier-Stark states along the lattice direction, the dominant terms are on-site and nearest-neighbor interactions. Since the characteristic $s$-wave interaction strength is much larger than $p$-wave interaction strength at ultracold temperatures ($\sim 100$ nK in our case), we include $p$-wave interaction only for on-site terms. All these approximations simplify Eq.~(\ref{eq:hamil}) into a large-spin Hamiltonian in a 1D lattice,
\begin{equation}
    \begin{gathered}
    H=H_{\mathrm{on-site}}+H_{\mathrm{off-site}}+H_{\mathrm{laser}},\\
    H_{\mathrm{on-site}}/\hbar=\sum_n\Big[J^{\perp}_0\mathbf{S}_n\cdot\mathbf{S}_n+\chi_0 S_n^zS_n^z+C_0 N_n S^z_n\Big],\\
    H_{\mathrm{off-site}}/\hbar=\sum_m\Big[J^{\perp}_1\mathbf{S}_n\cdot\mathbf{S}_{n+1}+\chi_1S_n^zS_{n+1}^z+D_1(S_n^xS_{n+1}^y-S_n^yS_{n+1}^x)\Big],\\
    H_{\mathrm{laser}}/\hbar=\sum_m\Big[-\delta S^z_n+\Omega_0 S^x_n\Big].\\
    \end{gathered}
    \label{eq:largespin}
\end{equation}
The interaction parameters for these collective spin operators are calculated by performing a thermal average over radial harmonic oscillator modes, 
\begin{equation}
    \begin{gathered}
    J^{\perp}_0=\eta_0(V_{eg}-U_{eg})/2, \quad \chi_0=\eta_0(V_{ee}+V_{gg}-2V_{eg})/2, \quad C_0=\eta_0(V_{ee}-V_{gg})/2, \\
    J^{\perp}_1=-\eta_1U_{eg}\cos\varphi, \quad \chi_1=-\eta_1U_{eg}(1-\cos\varphi),\quad D_1=-\eta_1U_{eg}\sin\varphi.\\
    \end{gathered}
\end{equation}
Here, $\eta_0$ and $\eta_1$ are dimensionless overlap integrals for on-site and nearest-neighbor interaction respectively [defined in Eq.~(\ref{eq:integral})], and the thermal average for $s$-wave ($U_{\alpha\beta}$) and $p$-wave ($V_{\alpha\beta}$) interaction strengths are
\begin{equation}
    U_{\alpha\beta}=\frac{8\pi\hbar a_{\alpha\beta}}{M}\frac{M\omega_R^2}{4\pi k_BT}\frac{k_L}{\sqrt{2\pi}}\bigg(\frac{V_0}{E_{rec}}\bigg)^{1/4}, \quad V_{\alpha\beta}=\frac{6\pi\hbar b^3_{\alpha\beta}}{M}\frac{1}{\pi}\bigg(\frac{M\omega_R}{\hbar}\bigg)^2\frac{k_L}{\sqrt{2\pi}}\bigg(\frac{V_0}{E_{rec}}\bigg)^{1/4}.
\end{equation}
The dependence of interaction parameters $\chi_0,\chi_1,C_0$ on lattice depth $V_0$ is shown in Fig.~\ref{supp_1}B.

\subsection{Density shift of the carrier transition}
As described in the main text, we measure the density shift of the carrier transition in Rabi spectroscopy. Note that the clock transition frequency is obtained by the average of two frequencies with the same excitation fraction on the positive and negative detuned side of the $\pi$-pulse Rabi spectrum, typically with an excitation fraction near $0.45$ (the maximum excitation fraction is near $0.9$). The density shift per atom
\begin{equation}
    \Delta\nu=\frac{\delta_{\mathrm{left}}+\delta_{\mathrm{right}}}{4\pi N_{\mathrm{loc}}},
\end{equation}
where $\delta_{\mathrm{left}}$ and $\delta_{\mathrm{right}}$ are the laser detuning from clock transition resonance for the excitation fraction we set on the positive and negative detuned side of the Rabi spectrum, and $N_{\mathrm{loc}}=\frac{1}{2L+1}\sum_{m=-L}^{L} N_{n+m}$ is the averaged atom number per site in a local region centered around site $n$. The local region is  $2L+1\sim 15$ lattice sites, corresponding to our $6$ $\mu$m imaging resolution.

To calculate the density shift, we apply a mean-field approximation to Eq.~(\ref{eq:largespin}),
\begin{equation}
    H_{\mathrm{MF}}/\hbar=\sum_n \mathbf{S}_n\cdot \mathbf{B}_n,
    \label{eq:mf1}
\end{equation}
where
\begin{equation}
\begin{gathered}
    B_n^x=\Omega_0+J^{\perp}_1(\langle S^x_{n-1}\rangle+\langle S^x_{n+1}\rangle)+D_1(\langle S^y_{n+1}\rangle-\langle S^y_{n-1}\rangle),\\
    B_n^y=J^{\perp}_1(\langle S^y_{n-1}\rangle+\langle S^y_{n+1}\rangle)-D_1(\langle S^x_{n+1}\rangle-\langle S^x_{n-1}\rangle),\\
    B_n^z=-\delta+2\chi_0\langle S^z_n\rangle+C_0N_n+J^{\perp}_1(\langle S^z_{n-1}\rangle+\langle S^z_{n+1}\rangle)+\chi_1(\langle S^z_{n-1}\rangle+\langle S^z_{n+1}\rangle).\\
    \end{gathered}
    \label{eq:b1}
\end{equation}
Note that we drop the $J^{\perp}_0$ term in Eq.~(\ref{eq:largespin}) because this term is a constant for any collective state at each lattice site. 

We further simplify Eq.~(\ref{eq:mf1}) by assuming all lattice sites share the same atom number $N_{\mathrm{loc}}$ in a local region (15 lattice sites). We calculate the spin dynamics in this local region by assuming  translationally invariant conditions $\langle S^{x,y,z}_n\rangle=\langle S^{x,y,z}\rangle$ to Eq.~(\ref{eq:b1}), where $\langle S^{x,y,z}\rangle=\frac{1}{2L+1}\sum_{m=-L}^{L} \langle S^{x,y,z}_{n+m}\rangle$. In this way, we have a homogeneous field on each site, $\mathbf{B}_n=\mathbf{B}$. The mean-field Hamiltonian becomes   
\begin{equation}
    H_{\mathrm{MF}}/\hbar=\sum_n \mathbf{S}_n\cdot \mathbf{B}^{\perp},
\end{equation}
where $\mathbf{B}^{\perp}\perp \langle\mathbf{S}\rangle$ is the perpendicular component of $\mathbf{B}$, with
\begin{equation}
    \mathbf{B}^{\perp}=\{\Omega_0,0,-\delta+2(\chi_0+\chi_1)\langle S^z\rangle+C_0N_{\mathrm{loc}}\}.
    \label{eq:b2}
\end{equation}
Dropping the parallel component of $\mathbf{B}$ because it does not contribute to the mean-field dynamics,   
\begin{equation}
    \frac{\mathrm{d}}{\mathrm{d}t}\langle\mathbf{S}_n\rangle=\mathbf{B}^{\perp}\times\langle\mathbf{S}_n\rangle.
\end{equation}
Using the mean-field equations above, we simulate the experimental protocol and obtain theoretical predictions for the density shift. In Rabi spectroscopy, we initialize all the atoms in the ground state ($\langle S^z_n\rangle=-N_{\mathrm{loc}}/2$) for $g\to e$ case, and all the atoms in the excited state ($\langle S^z_n\rangle=N_{\mathrm{loc}}/2$) for $e\to g$ case.

\begin{figure}[htp!]
\centering
\includegraphics[width=16 cm]{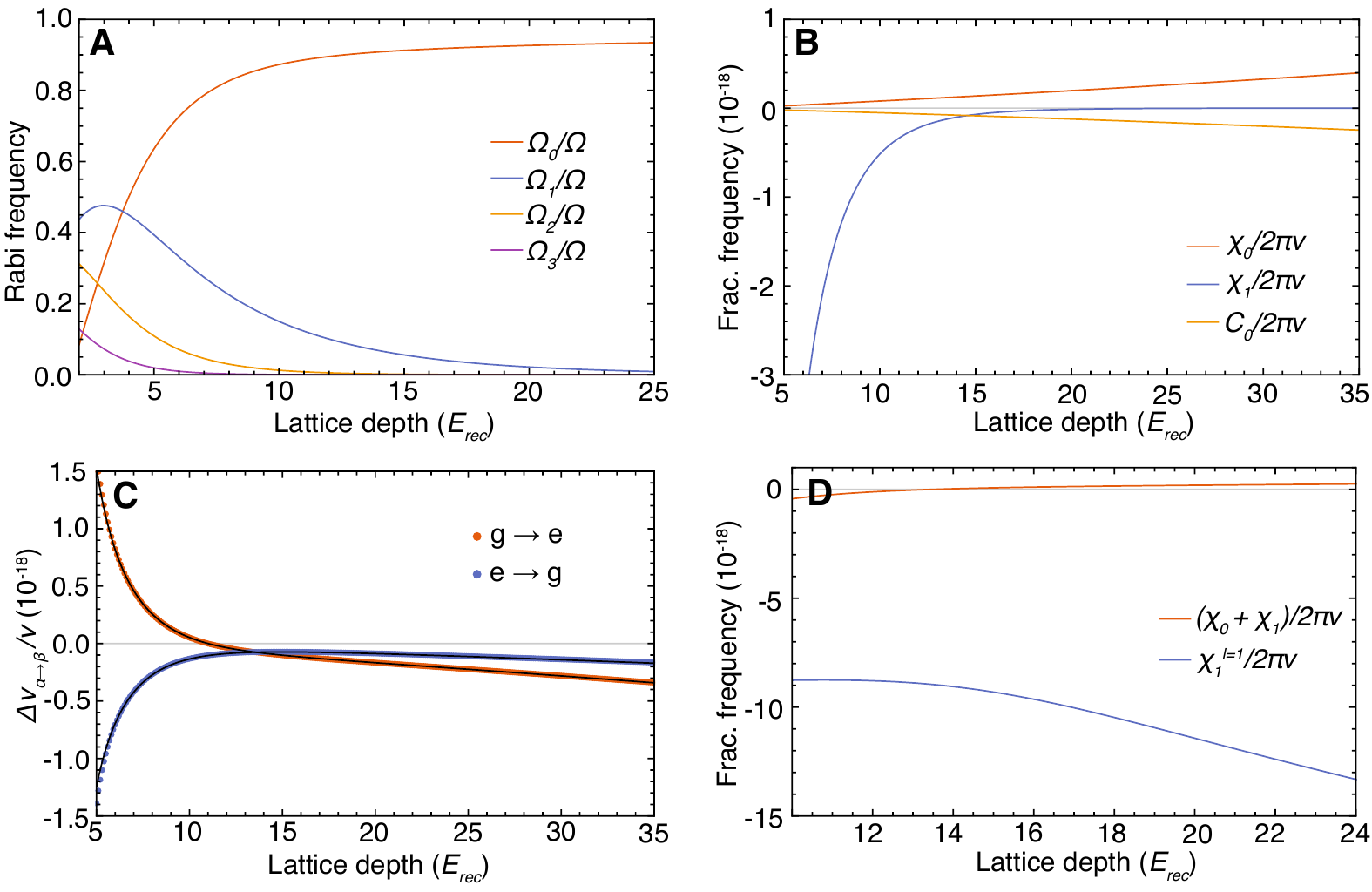}
\caption{\textbf{Spin Model Parameters.} 
(\textbf{A}) Rabi frequency for the  carrier transition ($\Omega_0$) and $l=1,2,3$ Wannier-Stark sidebands ($\Omega_1,\Omega_2,\Omega_3$) as a function of lattice depth. 
(\textbf{B}) Spin model parameters for the  carrier transition as a function of lattice depth. The radial temperature at each lattice depth  used  is the reported experimental value (See Fig. \ref{supp_0}.
(\textbf{C}) Theoretical predictions of the fractional frequency shift per atom (orange points for $g\rightarrow e$ case, blue points for $e\rightarrow g$ case) and numerical fits based on Eq.~(\ref{eq:fit}) shown as black lines. The fitting parameter used are  $\varsigma^z_{g \to e}=-0.12$ and $\varsigma^z_{e \to g}=0.095$. 
(\textbf{D}) Wannier-Stark sideband interaction parameter ($\chi_1^{l=1}$) compared to the carrier transition parameter ($\chi_0+\chi_1$), with the former significantly enhanced compared to the latter. In this case as well the radial temperature at each lattice depth is the reported experimental value \ref{temp}.
}
\label{supp_1}
\end{figure}

From Eq.~(\ref{eq:b2}), we obtain a simple expression for the density shift by setting it to be the value of $\delta$ at which $\mathbf{B}^{\perp}_z=0$:
\begin{equation}
\begin{gathered}
    \Delta\nu_{\alpha\to\beta}=\Delta\nu_{\alpha\to\beta}^s+\Delta\nu_{\alpha\to\beta}^p,\\
    2\pi\Delta\nu_{\alpha\to\beta}^p\approx2\chi_0\varsigma^z_{\alpha\to\beta}+C_0, \quad \quad  2\pi\Delta\nu_{\alpha\to\beta}^s\approx 2\chi_1\varsigma^z_{\alpha\to\beta}.
\end{gathered}
\label{eq:fit}
\end{equation}
Here, $\Delta\nu_{\alpha\to\beta}^{s,p}$ are the $s$-wave and $p$-wave contributions to the density shift, 
$\varsigma^z_{\alpha\to\beta}$ is a fitting parameter that accounts for the time evolution of $\langle S^z\rangle/N_{\mathrm{loc}}$ during the Rabi dynamics, which depends on the details of the Rabi drive such as the pulse area, excitation fraction, and initial conditions used in the experiment, $g\to e$ or $e \to g$. Based on our experimental condition in the carrier transition, we find $\varsigma^z_{g \to e}=-0.12$ and $\varsigma^z_{e \to g}=0.095$ [see Fig.~\ref{supp_1}(C)].
Note that $\Delta\nu_{\alpha\to\beta}^{p}$ are generated by on-site $p$-wave interactions, while $\Delta\nu_{\alpha\to\beta}^{s}$ are generated by nearest-neighbor $s$-wave interaction. This dependence allows us to control the density shift by adjusting the spatial extension of the Wannier-Stark states, which is  tunable by varying the lattice depth, the key idea to eliminating the density shift presented in the main text.

\subsection{Spin model for off-site Wannier-Stark transitions}
Apart from the carrier transition, we can also drive transitions to other Wannier-Stark states by using the clock laser to couple two different internal and motional states of an atom, $|\uparrow_{\mathbf{n}}\rangle\equiv|\tilde{e};n_X,n_Y,W_{n+l}\rangle$ and $|\downarrow_{\mathbf{n}}\rangle\equiv|\tilde{g};n_X,n_Y,W_n\rangle$, with $l=\pm1,\pm2,\cdots$.
Compared to the carrier transition, the subscript $\mathbf{n}$ labels different motional states for $|\uparrow_{\mathbf{n}}\rangle$ and $|\downarrow_{\mathbf{n}}\rangle$ states in off-site Wannier-Stark transitions. 
In this case, we expand the field operator $\psi_{\alpha}(\mathbf{r})$ in terms of single-particle eigenstates,
\begin{equation}
    \psi_{\tilde{e}}(\mathbf{R})=\sum_{\mathbf{n}}\phi_{n_X}(X)\phi_{n_Y}(Y)W_{n+l}(Z)c_{\mathbf{n}\uparrow}, \quad  \psi_{\tilde{g}}(\mathbf{R})=\sum_{\mathbf{n}}\phi_{n_X}(X)\phi_{n_Y}(Y)W_n(Z)c_{\mathbf{n}\downarrow}.
\end{equation}

Similar to the frozen-mode approximation used for the carrier transition, we treat each atom as a spin-$1/2$ system spanned by the $|\uparrow_{\mathbf{n}}\rangle$ and $|\downarrow_{\mathbf{n}}\rangle$ states defined for the specific Wannier-Stark states coupled by the laser, and rewrite the interaction Hamiltonian in terms of the corresponding spin operators,
\begin{equation}
    H_{\mathrm{int}}/\hbar=\sum_{\substack{\mathbf{n}\mathbf{m}\\(\mathbf{n}\neq\mathbf{m})}}\bigg[J^{\perp,l}_{\mathbf{n}\mathbf{m}}\mathbf{S}_{\mathbf{n}}\cdot\mathbf{S}_{\mathbf{m}}+\chi^{l}_{\mathbf{n}\mathbf{m}}S^z_{\mathbf{n}}S^z_{\mathbf{m}}+\frac{C_{\mathbf{n}\mathbf{m}}}{2}(S^z_{\mathbf{n}}N_{\mathbf{m}}+N_{\mathbf{n}}S^z_{\mathbf{m}})+\frac{K_{\mathbf{n}\mathbf{m}}^{l}}{2}(S^z_{\mathbf{n}}N_{\mathbf{m}}-N_{\mathbf{n}}S^z_{\mathbf{m}})\bigg],
\end{equation}
where
\begin{equation}
    \begin{gathered}
    J^{\perp,l}_{\mathbf{n}\mathbf{m}}=\eta_{|n-m|}^{\mathrm{ex},l}(V^{\mathrm{eg}}_{\mathbf{n}\mathbf{m}}-U^{\mathrm{eg}}_{\mathbf{n}\mathbf{m}})/2,\\
    \chi^{l}_{\mathbf{n}\mathbf{m}}=\eta_{|n-m|}(V^{ee}_{\mathbf{n}\mathbf{m}}+V^{gg}_{\mathbf{n}\mathbf{m}})/2-\eta_{|n-m|}^{\mathrm{dir},l}(V^{eg}_{\mathbf{n}\mathbf{m}}+U^{eg}_{\mathbf{n}\mathbf{m}})/2-\eta_{|n-m|}^{\mathrm{ex},l}(V^{\mathrm{eg}}_{\mathbf{n}\mathbf{m}}-U^{\mathrm{eg}}_{\mathbf{n}\mathbf{m}})/2,\\
    C_{\mathbf{n}\mathbf{m}}=\eta_{|n-m|}(V^{ee}_{\mathbf{n}\mathbf{m}}-V^{gg}_{\mathbf{n}\mathbf{m}})/2,\\
    K_{\mathbf{n}\mathbf{m}}^{l}=\eta^{\mathrm{diff},l}_{nm}(V^{eg}_{\mathbf{n}\mathbf{m}}+U^{eg}_{\mathbf{n}\mathbf{m}})/2.
    \end{gathered}
\end{equation}
Here, $\eta_{|n-m|}$, $U_{\mathbf{n}\mathbf{m}}^{\alpha\beta}$, $V_{\mathbf{n}\mathbf{m}}^{\alpha\beta}$ have the same definition as the ones used for the carrier transition [see Eq.~(\ref{eq:integral}) and Eq.~(\ref{eq:spd})], and the definitions for the extra dimensionless overlap intergrals are
\begin{equation}
\begin{gathered}
    \eta_{|n-m|}^{\mathrm{dir},l}=\frac{1}{2}(\eta_{|n-m+l|}+\eta_{|n-m-l|}), \\
    \eta_{nm}^{\mathrm{diff},l}=\frac{1}{2}(\eta_{|n-m+l|}-\eta_{|n-m-l|}), \\
    \eta_{|n-m|}^{\mathrm{ex},l}=\frac{\lambda_L}{\sqrt{2\pi}}\bigg(\frac{V_0}{E_{rec}}\bigg)^{-1/4}\int\mathrm{d}Z\,W_n(Z)W_m(Z)W_{n+l}(Z)W_{m+l}(Z).
\end{gathered}
\end{equation}

Note that the Rabi frequency for the Wannier-Stark sidebands experiences the same spin-orbit coupling phase as the carrier transition. We use the gauge transformation as in  the carrier transition to redefine the spin operators [see Eq.~(\ref{eq:gauge})], and the effective Hamiltonian in the gauged frame becomes
\begin{equation}
    \begin{aligned}
    H/\hbar&=\sum_{\substack{\mathbf{n}\mathbf{m}\\(\mathbf{n}\neq\mathbf{m})}}\bigg[\tilde{J}^{\perp,l}_{\mathbf{n}\mathbf{m}}\tilde{\mathbf{S}}_{\mathbf{n}}\cdot\tilde{\mathbf{S}}_{\mathbf{m}}+\tilde{\chi}^{l}_{\mathbf{n}\mathbf{m}}\tilde{S}^z_{\mathbf{n}}\tilde{S}^z_{\mathbf{m}}+D^{l}_{\mathbf{n}\mathbf{m}}(\tilde{S}^x_{\mathbf{n}}\tilde{S}^{y}_{\mathbf{m}}-\tilde{S}^y_{\mathbf{n}}\tilde{S}^{x}_{\mathbf{m}})+\frac{C_{\mathbf{n}\mathbf{m}}}{2}(\tilde{S}^z_{\mathbf{n}}\tilde{N}_{\mathbf{m}}+\tilde{N}_{\mathbf{n}}\tilde{S}^z_{\mathbf{m}})\\
    &\frac{K_{\mathbf{n}\mathbf{m}}^{l}}{2}(\tilde{S}^z_{\mathbf{n}}\tilde{N}_{\mathbf{m}}-\tilde{N}_{\mathbf{n}}\tilde{S}^z_{\mathbf{m}})\bigg]-\hbar\delta_l\sum_{\mathbf{n}}\tilde{S}^z_{\mathbf{n}}+\hbar\Omega_l\sum_{\mathbf{n}}\tilde{S}^x_{\mathbf{n}},
    \end{aligned}
\end{equation}
where $\tilde{J}^{\perp,l}_{\mathbf{n}\mathbf{m}}=\cos[(n-m)\varphi]J^{\perp}_{\mathbf{n}\mathbf{m}}$, $\tilde{\chi}^{l}_{\mathbf{n}\mathbf{m}}=\chi^{l}_{\mathbf{n}\mathbf{m}}+J^{\perp,l}_{\mathbf{n}\mathbf{m}}-\tilde{J}^{\perp,l}_{\mathbf{n}\mathbf{m}}$, $D^{l}_{\mathbf{n}\mathbf{m}}=-\sin[(n-m)\varphi]J^{\perp,l}_{\mathbf{n}\mathbf{m}}$, $\delta_l=\delta-lMga_L/\hbar$, and 
\begin{equation}
    \Omega_l=\Omega\cdot \mathcal{C}\mathcal{J}_l\bigg(\frac{4J_0}{Mga_L}\sin(\varphi/2)\bigg).
\end{equation}
The dependence of $\Omega_l$ ($l=1,2,3$) on lattice depth $V_0$ is shown in Fig.~\ref{supp_1}(A).

In the following discussions, we focus on the $l=1$ Wannier-Stark transition. Following the same procedure we used for the carrier transition, we can express the Hamiltonian dynamics in terms of collective spin operators, $S^{x,y,z}_n=\sum_{n_xn_y}\tilde{S}_{\mathbf{n}}^{x,y,z}$, $N_n=\sum_{n_xn_y}\tilde{N}_{\mathbf{n}}$. 
Recall that for the carrier transition we discussed in previous sections, on-site $s$-wave interactions only gave rise to a constant term ($J_0^{\perp}$ term) which does not play any role in the  mean-field dynamics. The dominant interaction comes from on-site $p$-wave interactions and nearest-neighbor $s$-wave interactions. However, in the case of the $l=1$ site-changing Wannier-Stark transition, a ground state atom in $|W_n\rangle$ acquires a non-zero admixture of the excited state in $|W_{n+1}\rangle$. This component can interact with a ground state atom in $|W_{n+1}\rangle$ via $s$-wave interactions. Since the on-site $s$-wave interactions play a significant role in this case, we drop all the interaction terms smaller than such  on-site $s$-wave interactions. We can also drop the $K^{l}_{\mathbf{n}\mathbf{m}}$ term due to the uniform atom population for lattice sites in a local regime. These approximations lead to the following large-spin Hamiltonian in a 1D lattice, 
\begin{equation}
    H/\hbar=\sum_n\Big[\chi_1^{l=1}S_n^zS_{n+1}^z-\delta_1 S^z_n+\Omega_1 S^x_n\Big],
    \label{eq:largeside}
\end{equation}
where
\begin{equation}
    \chi_1^{l=1}=-\eta_0U_{eg}/2.
\end{equation}
In Fig.~\ref{supp_1}(D), we compare $\chi_1^{l=1}$ with its counterpart $\chi_0+\chi_1$ in the carrier transition. It is clear that the interaction is significantly enhanced due to site-changing Wannier-Stark transitions.

\subsection{Dynamical phase transition}
In the main text we presented theoretical and experimental results on the ferromagnetic to paramagentic dynamical phase transition (DPT) when we address the $l = 1$  Wannier-Stark transition. Given that Eq.~(\ref{eq:largeside}) is a large-spin Hamiltonian, its dynamical phase diagram is well captured by a mean-field approximation. Similar to the carrier transition, we apply the translationally invariant condition $\langle S^{x,y,z}_n\rangle=\langle S^{x,y,z}\rangle$ in a local regime (15 lattice sites), where $\langle S^{x,y,z}\rangle=\frac{1}{2L+1}\sum_{m=-L}^{L}\langle S^{x,y,z}_{n+m}\rangle$. This leads to the following mean-field Hamiltonian,
\begin{equation}
    H_{\mathrm{MF}}/\hbar=\sum_n\mathbf{S}_n\cdot\mathbf{B},
    \label{HamDPT}
\end{equation}
where
\begin{equation}
    \mathbf{B}=\{\Omega_1,0,-\delta_1+2\chi_1^{l=1}\langle S^z\rangle\}.
\end{equation}
Writing mean-field equations can be written in terms of normalized expectation value of collective spin operators on a single site $s^{x,y,z}=2\langle S^{x,y,z}\rangle/N_{\mathrm{loc}}$,
\begin{equation}
    \begin{gathered}
    \frac{\mathrm{d}}{\mathrm{d}t}s^x=-N_{\mathrm{loc}}\chi_1^{l=1} s^zs^y+\delta_1 s^y,\\
    \frac{\mathrm{d}}{\mathrm{d}t}s^y=N_{\mathrm{loc}}\chi_1^{l=1} s^zs^x-\delta_1 s^x-\Omega_1 s^z,\\
    \frac{\mathrm{d}}{\mathrm{d}t}s^z=\Omega_1 s^y.\\
    \end{gathered}
    \label{eq:dpt1}
\end{equation}
Note that Eq.~(\ref{eq:dpt1}) takes the same form as the mean-field equations obtained in \cite{Muniz2020,Chu2020}, which predicted a DPT between ferromagnetic and paramagnetic phases. 

\begin{figure}[htp!]
\centering
\includegraphics[width=16 cm]{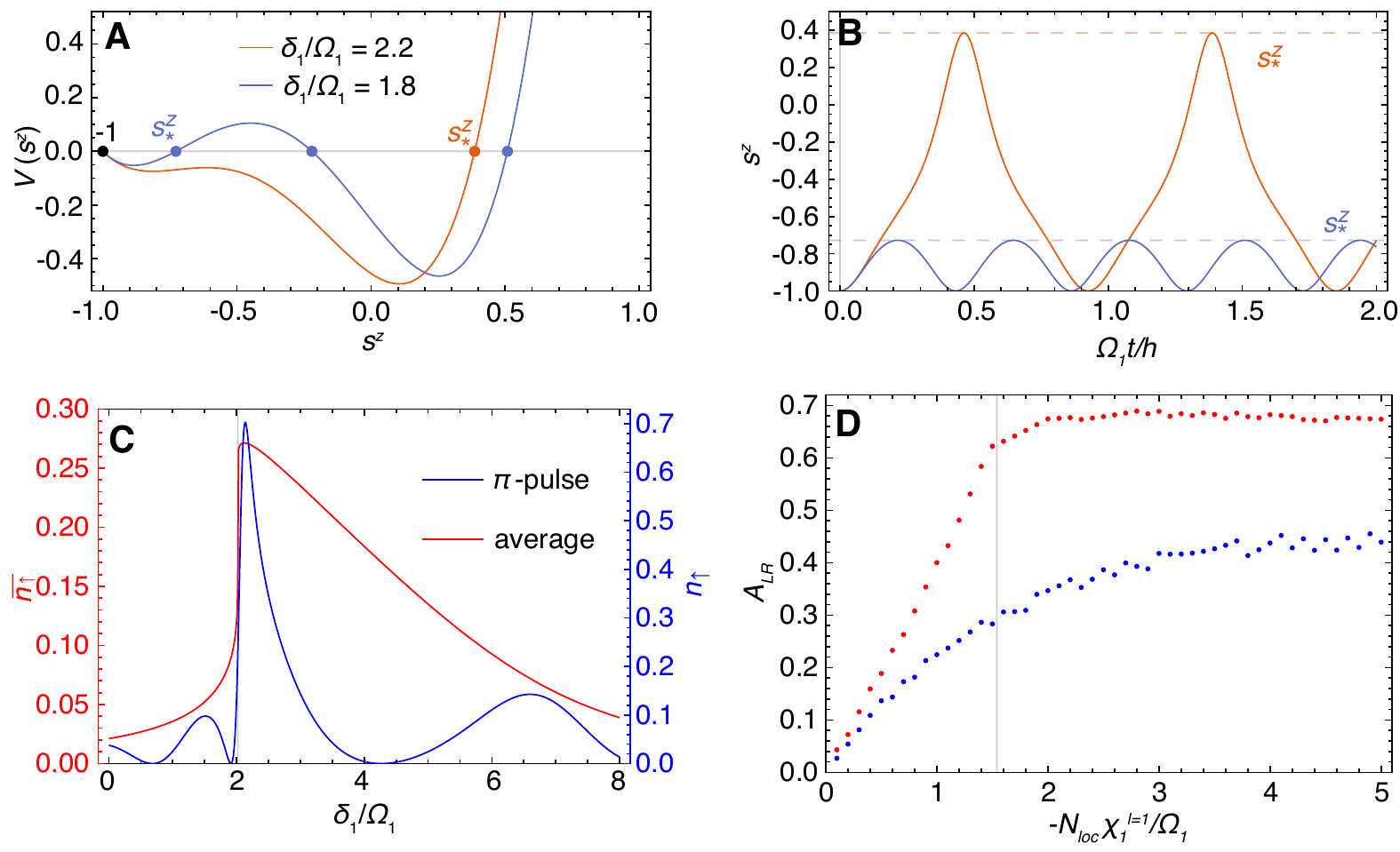}
\caption{\textbf{Dynamical Phase Transition.} 
(\textbf{A}) The effective potential $V(s^z)$ with $N_{\mathrm{loc}}\chi_1^{l=1}/\Omega_1=-5$. In the case of $\delta_1/\Omega_1=2.2$, $V(s^z)$ has two real roots; In the case of $\delta_1/\Omega_1=1.8$, $V(s^z)$ has four real roots. The nearest turnover point is labelled by $s^z_{*}$, and the jump of $s^z_{*}$ indicates the DPT.
(\textbf{B}) The mean-field dynamics of our model with $N_{\mathrm{loc}}\chi_1^{l=1}/\Omega_1=-5$ and $\delta_1/\Omega_1=2.2,1.8$, which shows a sharp change of mean-field dynamical behavior. The choice of color for the lines is the same as (\textbf{A}).
(\textbf{C}) The long-time average excitation fraction $\overline{n_{\uparrow}}$ (red line) and the Rabi lineshape after a $\pi$-pulse (blue line) with $N_{\mathrm{loc}}\chi_1^{l=1}/\Omega_1=-5$. The critical point (marked by gray line) that  separates the ferromagnetic phase (left) and paramagnetic phase (right) is captured by the maximum derivative in both of the curves. 
(\textbf{D}) Asymmetry of the long-time averaged excitation fraction and Rabi lineshape in (\textbf{C}) with the same choice of color. The gray line separates the crossover regime (left) and DPT regime (right).
}
\label{supp_2}
\end{figure}

In general terms, a DPT is characterized by the existence of a critical point separating phases with distinct dynamical properties in many-body systems after a sudden quench. The analog of thermodynamic order parameters is found in long-time average observables, which have a non-analytic dependence on system parameters. We initialize all the atoms in the $|\downarrow\rangle$ state, the ground state of our model when $\delta_1\rightarrow-\infty$, and then perform a sudden quench of the longitudinal field to its final value $\delta_1$. The DPT is signaled by a sharp change in behavior of the long-time average excitation fraction $\overline{n_{\uparrow}}=(\overline{s^z}+1)/2$, where $\overline{s^z}=\lim_{T\rightarrow\infty}\frac{1}{T}\int_0^Ts^z(t)\mathrm{d}t$.
In the dynamical ferromagnetic phase, $\overline{n_{\uparrow}}\approx 0$ persists even when the final longitudinal field $\delta_1$ is varied. In the dynamical paramagnetic phase, $\overline{n_{\uparrow}}$ dynamically adjusts itself following the change of final longitudinal field $\delta_1$ [See Fig.~\ref{supp_2}(C)].

In the following, we analyze the critical points for the DPT in our system based on the procedure described in \cite{Muniz2020,Chu2020}.
Using energy conservation in $H_{\mathrm{MF}}$ for an initial state with $s^z=-1, \,s^x=s^y=0$, 
\begin{equation}
    \frac{N_{\mathrm{loc}}\chi_1^{l=1}}{2}s^zs^z-\delta_1s^z+\Omega_1 s_x=\frac{N_{\mathrm{loc}}\chi_1^{l=1}}{2}+\delta_1,
    \label{eq:dpt2}
\end{equation}
as well as the identity,
\begin{equation}
    (s^x)^2+(s^y)^2+(s^z)^2=1.
    \label{eq:dpt3}
\end{equation}
In the large-$N_{\mathrm{loc}}$ limit, we can eliminate $s^x$ and $s^y$, and obtain the following differential equation for $s^z$,
\begin{equation}
    \frac{1}{2}\bigg(\frac{\mathrm{d}}{\mathrm{d}t}s^z\bigg)^2+V(s^z)=0,
    \label{eq:dpt4}
\end{equation}
where
\begin{equation}
    \begin{aligned}
    V(s^z)=(s^z+1)&\bigg\{\frac{(N_{\mathrm{loc}}\chi_1^{l=1})^2}{8}(s^z)^3-\bigg[\frac{(N_{\mathrm{loc}}\chi_1^{l=1})^2}{8}+\frac{N_{\mathrm{loc}}\chi_1^{l=1}\delta_1}{2}\bigg](s^z)^2\\
    &+\bigg[\frac{\delta_1^2+\Omega_1^2}{2}-\frac{(N_{\mathrm{loc}}\chi_1^{l=1})^2}{8}\bigg]s^z+\bigg[\frac{\delta_1^2-\Omega_1^2}{2}+\frac{N_{\mathrm{loc}}\chi_1^{l=1}\delta_1}{2}+\frac{(N_{\mathrm{loc}}\chi_1^{l=1})^2}{8}\bigg]\bigg\}.
    \end{aligned}
\end{equation}
Our experimental conditions lie in the parameter regime where $N_{\mathrm{loc}}\chi_1^{l=1}<0$ with a fixed positive $\Omega_1$.

We interpret Eq.~(\ref{eq:dpt4}) as the Hamiltonian of a classical particle with position $s^z$ moving in the effective potential $V(s^z)$, which is shown in Fig.~\ref{supp_2}(A). The condition $V(s^z)=0$ determines the physical turnover points of $s^z$. Since $V(-1)=0$, $V'(-1)=-1$, $V(1)=2\delta_1^2$, this effective potential has at least two real roots in $[-1,1]$. The dynamics of $s^z$ can be understood as the oscillations between $-1$ and the nearest turnover point $s^z_{*}$ [see Fig.~\ref{supp_2}(B)]. Suppose we start from a $V(s^z)$ with two real roots, and continuously tune the parameters of $V(s^z)$ so that two new real roots appear in between. Then a jump of the nearest turnover point $s^z_{*}$ should occur in this process. This abrupt change in behavior is what sets the dynamical phase transition [see Fig.~\ref{supp_2}(A,B)].

To count the number of roots in $V(s^z)$, we factor out the known root $s^z=-1$, and then consider the discriminant $\Delta=18abcd-4b^3d+b^2c^2-4ac^3-27a^2d^2$ of cubic equation $ax^3+bx^2+cx+d=0$. If $\Delta>0$, the cubic equation has three distinct real roots; if $\Delta<0$, the cubic equation has one real root. So $\Delta=0$ sets the critical points of the DPT, presented as the black solid line in Fig.~4(C) of the main text. As shown in Fig.~\ref{supp_2}(C), the critical points can be captured by the divergence of the first derivative of $\overline{n_{\uparrow}}$. Similar to \cite{Muniz2020,Chu2020}, our experiment measures the excitation fraction at a finite time (after a $\pi$ pulse) instead of the long-time averaged excitation fraction. Although the derivative does not diverge in experiment, the maximum derivative can  still be used  to capture the critical point as shown in  Fig.~\ref{supp_2}(C)]. In Fig.~4(C) of the main text, we construct the phase boundary of the DPT with the maximum derivative of the experimental Rabi lineshapes. We find that the many-body decoherence discussed in the next section has negligible effect on the position of the critical points, nevertheless it obscures  the   sharp features  at the DPT  expected from    Eq. (\ref{HamDPT}).        

Moreover, based on the existence of real roots in equation $\Delta=0$, we can also differentiate  the DPT regime  ($N_{\mathrm{loc}}\chi_1^{l=1}/\Omega_1<-8\sqrt{3}/9$) dominated by interactions and the smooth crossover regime ($-8\sqrt{3}/9<N_{\mathrm{loc}}\chi_1^{l=1}/\Omega_1<0$) dominated by single-particle Rabi flopping where no DPT takes place. Based on our experimental condition ($22E_{rec}$, $190$nK and $2.3$s $\pi$-pulse), the boundary of these two regimes $N_{\mathrm{loc}}\chi_1^{l=1}/\Omega_1=-8\sqrt{3}/9$ is equivalent to $N_{\mathrm{loc}}=63.4$, indicated by the black dashed line in Fig.~4(C) of the main text. These two regimes can also be determined by the asymmetry of the long-time averaged excitation fraction or Rabi lineshape,  defined as $A_{LR}=(n_R-n_L)/(n_R+n_L)$. Here, $n_R=\int_{\delta_{max}}^{\delta_{max}+f} n_{\uparrow}(\delta)\mathrm{d}\delta$, $n_L=\int_{\delta_{max}-f}^{\delta_{max}} n_{\uparrow}(\delta)\mathrm{d}\delta$, where $\delta_{max}$ is the detuning at which  the peak value of $n_{\uparrow}$ is reached, and we choose $f$ to cover almost the entire frequency range of non-vanishing $n_{\uparrow}$. In Fig.~\ref{supp_2}(D), we compare the $A_{LR}$ obtained from the long-time averaged excitation fraction and the the one obtained from the Rabi lineshape after a $\pi$ pulse. In both cases, the asymmetry $A_{LR}$ becomes more pronounced as the atom number increases in the crossover regime, while $A_{LR}$ saturates near the maximum value in the DPT regime. Note that the many-body decoherence discussed in the next section generally reduces the asymmetry. Nevertheless  the saturation behavior observed  in the DPT regime is maintained. For convenience, in Fig.~4(C) of the main text we normalize the maximum value of $A_{LR}$ obtained from experimental lineshapes to $1$.

\section{Additional Experimental Details}

\subsection{Sample preparation}\label{temp}
\begin{figure}[thp!]
\centering
\includegraphics[width=9 cm]{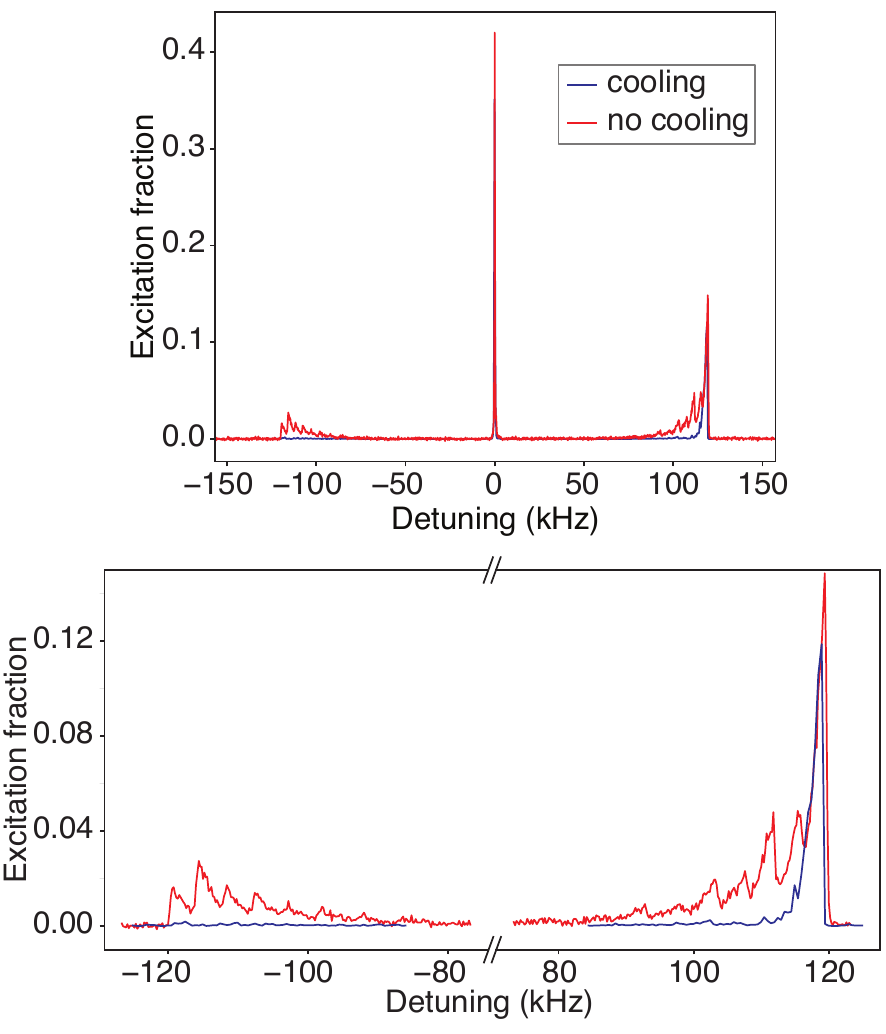}
\caption{\textbf{Sideband Spectroscopy.} Axial sideband spectroscopy with a radially cooled sample at $300$ $E_{rec}$. 
We scan the detuning of the clock laser from the carrier transition using high intensity to probe the axial mode filling. 
The red trace was measured without axial sideband cooling.
The blue trace was measured after implementing axial sideband cooling, illustrating near perfect sample preparation in the lowest motional band.
The different axial transitions are apparent in the axial sidebands enlarged in the lower plot.
}
\label{supp_axial}
\end{figure}

We prepare a nuclear spin polarized, cold sample at a high lattice depth of $300$ $E_{rec}$ and then adiabatically reduce the lattice depth.
To ensure the atoms are in the lowest motional band along the lattice (axial) direction, we utilize standard sideband cooling techniques and probe the mode filling using sideband spectroscopy, shown in Fig. \ref{supp_axial}. 
Without axial cooling, the red trace, the sample thermally populates many axial modes.
Due to the anharmonicity of the trapping potential, transitions between different oscillator levels are resolvable as distinct peaks in both the positively detuned blue and negatively detuned red sidebands.
With sideband cooling, the red sideband is entirely eliminated and the blue sideband shows only a single axial mode, indicating sample preparation in the axial ground motional state.
This purity is confirmed after lattice ramping to the operational depth.

As in Ref. \cite{blatt2009}, we fit the sideband spectra to extract a lattice depth.
At very low lattice intensities, this fitting technique is no longer a reliable method to determine the lattice depth.
Instead, we use the transmitted lattice power and depth fits at higher intensities to calculate the lattice depths at our operational points. 

\begin{figure}[htp!]
\centering
\includegraphics[width= 7 cm]{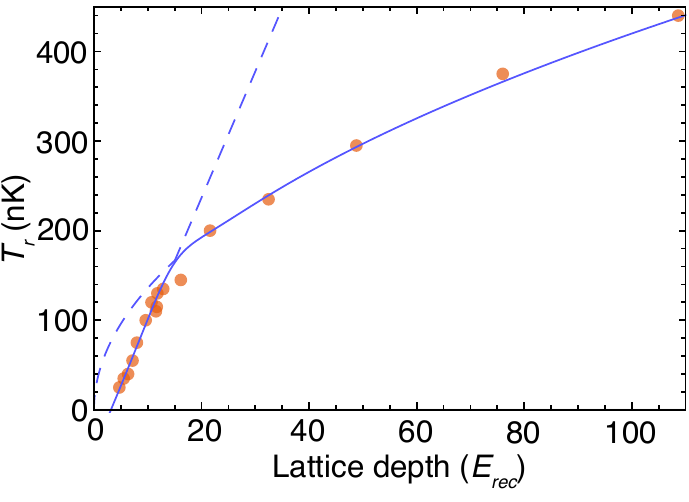}
\caption{\textbf{Radial Temperature.} Radial temperature $T_r$ measured over a range of operational lattice depths $V_0$ in units of lattice photon recoil energies, $E_{rec}$.
The red points indicate experimental data, and the solid blue lines show the piecewise fit of equation \ref{eq_radial}.
}
\label{supp_0}
\end{figure}
In addition to axial sideband cooling, we use field free Doppler cooling to reduce the radial temperature.
We measure the radial temperature $T_r$ of the ensemble before each density shift measurement. 
Driving the narrow clock transition with a beam oriented perpendicular to the lattice direction, we scan the clock laser frequency and measure the excitation fraction of the sample. 
This excitation profile is fit with a radial Doppler absorption profile to extract $T_r$.

Our camera based imaging spectroscopy technique provides a spatial map of the temperature throughout the mm length cloud. We observe temperature variations of up to $10$ nK over the entire sample, with the mean temperature presented in Fig. \ref{supp_0}.
The temperature over the range of operational lattice depths $V_0$ is well described by:
\begin{equation}
    T_r(nK)=\begin{cases}
       -45.2+14.1V_0/E_{rec} \quad (V_0<15E_{rec}),\\
       42\sqrt{V_0/E_{rec}} \quad (V_0>15E_{rec}).\\
    \end{cases}
    \label{eq_radial}
\end{equation}
For $V_0 > 15$ $E_{rec}$, the trend of $T_r$ matches that expected from an adiabatic lowering of the trap depth. At sufficiently low lattice depths, the non harmonic radial trap behavior leads to a deviation from adiabatic temperatures. For the lowest values of $T_r$ approaching 20 nK, our Doppler spectroscopy technique is also reaching its limit of reliability. 

\subsection{Coherence time of off-site Wannier-Stark transitions}
\begin{figure}[htp!]
\centering
\includegraphics[width=7 cm]{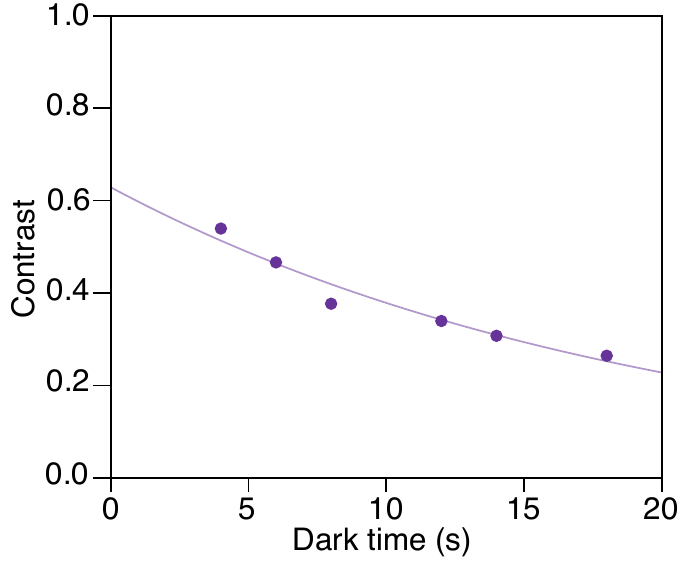}
\caption{\textbf{Coherence Time.} To determine the coherence time of a site-changing Wannier-Stark transition, we use a Ramsey sequence with a randomly sampled phase for the second rotation. 
As in Ref. \cite{bothwell2021}, randomly sampling a phase for a given dark time traces out an ellipse.
We fit the ellipse to extract a contrast measurement, reported here as purple dots.
The contrast decay is fit with a single exponential with decay time $\tau = 20(1)$ s. 
}
\label{supp_coherence}
\end{figure}

We utilize a similar method as in Ref.~\cite{bothwell2021} to extract a coherence time on the $|g \, ; \, W_n\rangle \rightarrow |e \, ; \, W_{n\pm 1} \rangle$ transition.  Using a Ramsey sequence with a randomly sampled phase for the second pulse, we fit the contrast decay between two spatially resolved regions of the sample as a function of dark time, see Fig.~\ref{supp_coherence}. 

Due to very strong $s$-wave interactions on this transition, we observe a significant dependence on local density.
While this dependence merits further study, here we operate the system in a relatively low density regime compared to our density shift measurements to determine a near optimal coherence. Fitting a single exponential time decay to the contrast, we measure an atomic coherence time of $20(1)$ s.
The regions of study are selected by fitting a Gaussian to the atomic distribution and selecting two regions from the center to $1.5$ times the Gaussian RMS width.
Greater coherence time was observed in regions with lower density. Hence what we report here represents the lower bound of atomic coherence for the off-site Wannier-Stark drive. There is no reason to expect its coherence to be less than that of the carrier transition, if one can cleanly separate the interaction effects.  

\subsection{Density shift measurement}

\begin{figure}[htp!]
\centering
\includegraphics[width=16 cm]{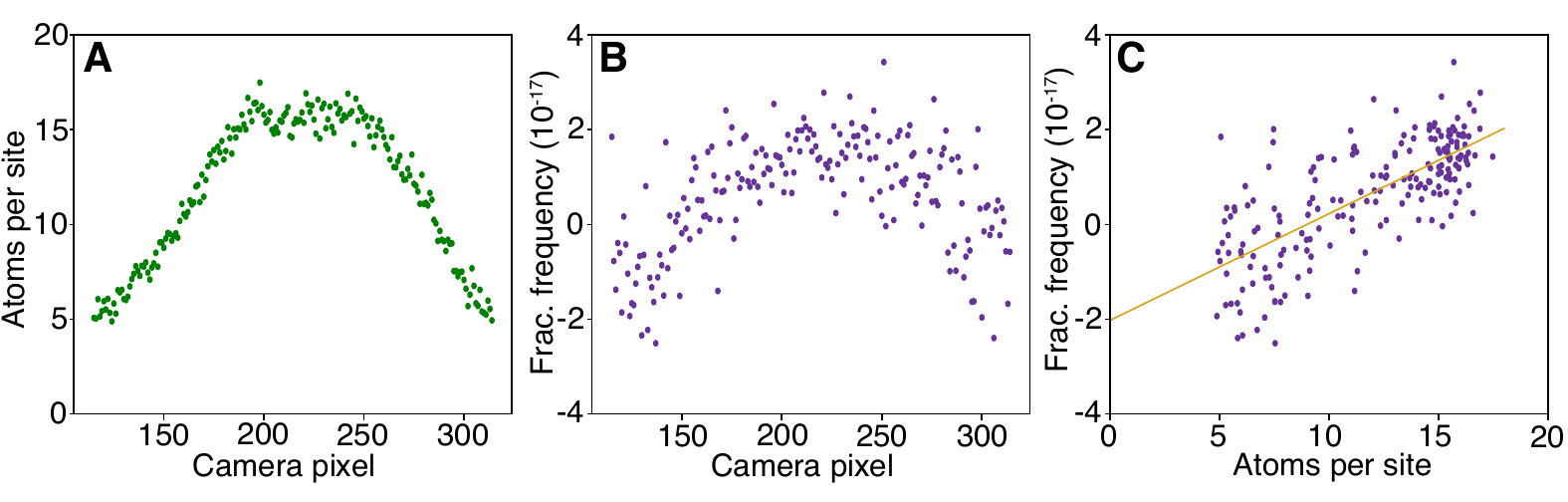}
\caption{\textbf{Measuring the Density Shift.} 
For each experiment cycle we collect a set of images that allow us to locally determine the excitation fraction and atom number.
To optimize the signal, we select a region of interest near the peak density as in Ref. \cite{bothwell2021}.
With a four point probing scheme, we measure the average atom number at each pixel (\textbf{A}) and construct the bare frequency (\textbf{B}).
The absolute frequency is arbitrary.
(\textbf{C}) The frequency as a function of atoms per site.
The purple points are data and the gold line is a linear fit, the slope of which is the density shift coefficient for this single four point measurement cycle.
}
\label{supp_density_fit}
\end{figure}

As described in the main text, we use extended `clock locks' to measure the spatially dependent average density shift. For each experimental cycle consisting of four Rabi probes, we construct a frequency map throughout the sample using in situ imaging, as shown from an example measurement in Fig.~\ref{supp_density_fit}.
The total number of counts at each pixel is proportional to the number of atoms (Fig.~\ref{supp_density_fit}A).
The corresponding atom number is calibrated using the standard quantum projection noise techniques~\cite{itano1993}. The four laser frequency lock points probe opposite sign $m_F$ transitions. From the mean frequency of these two transitions, we find the transition frequency at each pixel, as shown in Fig.~\ref{supp_density_fit}B.
Finally, as illustrated in Fig.~\ref{supp_density_fit}C, we fit this frequency as a function of atom number with a linear model, weighting by the atom number to account for quantum projection noise. 
The slope of this fit is the density shift coefficient for one experimental cycle.
For each lattice dpeth, the density shift coefficient reported in the main text is the mean of all the coefficients measured over an extended clock lock sequence, with the uncertainty arising from an Allan deviation of these coefficients at $1/6$ the measurement time. 

\section{Experiment - Theory Comparisons}
\subsection{Many-body decoherence in off-site Wannier-Stark transitions}

\begin{figure}[htp!]
\centering
\includegraphics[width=12 cm]{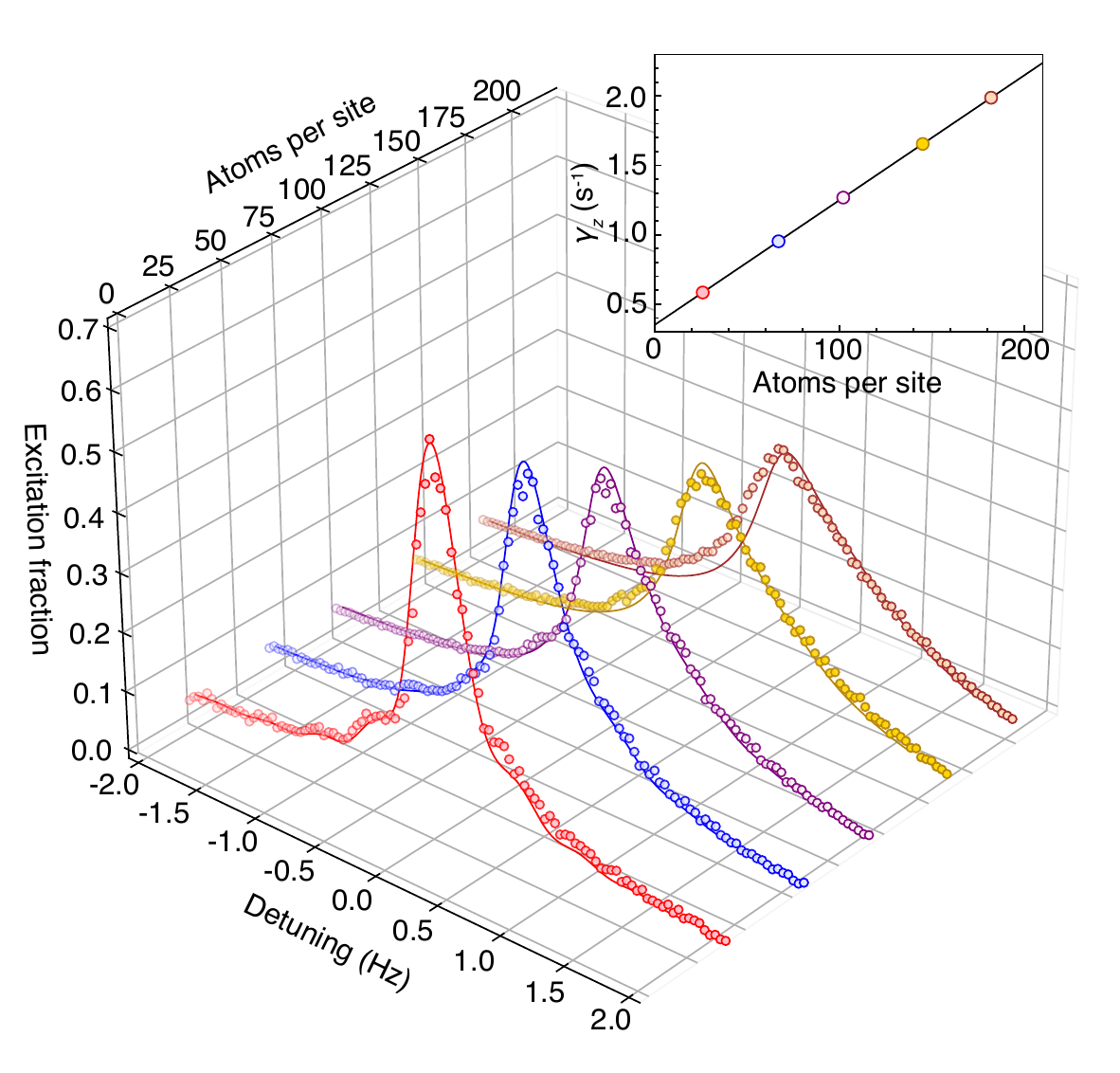}
\caption{\textbf{Many-body Decoherence.} Rabi lineshapes for the $l=1$ Wannier-Stark $l=1$ transition and corresponding  theoretical fits at different $N_{\mathrm{loc}}$. The dephasing rate $\gamma_z$ is the only fitting parameter, which is shown in the inset using the same color as the Rabi lineshapes. The linear dependence of $\gamma_z$ on atom number per site ($\gamma_z=0.35+0.009N_{\mathrm{loc}}$) confirms  that the dephasing effect is generated by mode-changing collisions. 
}
\label{supp_3}
\end{figure}

Our theoretical model is based on the frozen-mode approximation, which restricts the  accessible Hilbert space of each atom into a spin-$1/2$  degree of freedom spanned by the  $|\uparrow_{\mathbf{n}}\rangle$ and $|\downarrow_{\mathbf{n}}\rangle$ states. Our spin model is valid in the collisionless regime, breaking down at long times or at high enough densities where mode relaxation is not negligible. Since the interaction strength is significantly enhanced when interrogating site-changing WS transitions, as  discussed in previous sections, the mode relaxation rate is expected to be more significant. We take into account  the mode-changing collisions phenomenologically by adding a density-dependent dephasing term ($\gamma_z$) into our mean-field equations for the $l=1$ Wannier-Stark sideband [see Eq.~(\ref{eq:dpt1})],
\begin{equation}
    \begin{gathered}
    \frac{\mathrm{d}}{\mathrm{d}t}s^x=-N_{\mathrm{loc}}\chi_1^{l=1} s^zs^y+\delta_1 s^y-\gamma_z s^x,\\
    \frac{\mathrm{d}}{\mathrm{d}t}s^y=N_{\mathrm{loc}}\chi_1^{l=1} s^zs^x-\delta_1 s^x-\Omega_1 s^z-\gamma_z s^y,\\
    \frac{\mathrm{d}}{\mathrm{d}t}s^z=\Omega_1 s^y,\\
    \end{gathered}
\end{equation}
We use $\gamma_z$  as a fitting parameter  and find it has  a linear dependence on $N_{\mathrm{loc}}$ as expected from mode changing decoherence. In Fig.~\ref{supp_3}, we compare our theoretical predictions with the Rabi lineshapes observed in experiment at different $N_{\mathrm{loc}}$, with good agreement by setting the dephasing rate $\gamma_z=0.35+0.009N_{\mathrm{loc}}$. Small deviations are observed at  the highest densities approaching  200 atoms per site.  
 
\subsection{Scattering parameters}
In Ref.~\cite{martin2013,zhang2014spectroscopic}, the relation  between the $p$-wave interaction and $p$-wave scattering length was missing a factor of 1/2, with the correct coefficient being $3\pi\hbar^2b^3_{\alpha\beta}/2M$. Using the past notation, in Ref.~\cite{zhang2014spectroscopic} the $p$-wave scattering lengths were found to be:  $\tilde{b}_{eg}^{+}=(-169\pm 23)a_0$, $\tilde{b}_{ee}=(-119\pm 18)a_0$. These values can be corrected by solving:  $b_{ee}^3-b_{gg}^3=2(\tilde{b}_{ee}^3-b_{gg}^3)$ and $(b_{eg}^{+})^3-b_{gg}^3=2((\tilde{b}_{eg}^{+})^3-b_{gg}^3)$, which gives $b_{eg}^{+}=(-215.9\pm 28.2)a_0$ and $b_{ee}=(-155.8\pm 21.1)a_0$. For $p$-wave inelastic scattering length $\beta_{ee}$, one can multiply the factor $2^{1/3}$ to the value in Ref.~\cite{zhang2014spectroscopic}, which gives $\beta_{ee}=(152.5\pm 16.4)a_0$. Using these corrected values of the $p$-wave parameters, combined with the measured $s$-wave scattering lengths in Ref.~\cite{goban2018emergence}, as well as the universal relation between the complex $s$-wave scattering length $A=a-i\alpha$ and the complex $p$-wave scattering volume $B^3=b^3-i\beta^3$ for a single van der Waals potential \cite{zhang2014spectroscopic,Idziaszek2010}, one can finally obtain Table 1 that includes the updated  $s$-wave and $p$-wave scattering lengths that are used in this work.
\begin{table}[h!]
  \renewcommand\arraystretch{1.5}
  \begin{center}
    \caption{$^{87}$Sr $s$-wave and $p$-wave scattering lengths in Bohr radius ($a_0$)}
    \begin{tabular}{ c | c | c }

      \hline 
      Channel & $s$-wave & $p$-wave\\
      \hline 
      $gg$ & $96.2\pm 0.1$ & $74.5\pm 0.3$\\
      $eg^{+}$ & $161.3\pm 2.5$ & $-215.9\pm 28.2$\\
      $eg^{-}$ & $69.1\pm 0.9$ & $-41.3\pm 2.7$\\
      $ee$ (elastic) & $176.3\pm 9.5$ & $-155.8\pm 21.1$\\
      $ee$ (inelastic) & $17.3^{+14}_{-8}$ & $152.5\pm 16.4$\\
      \hline
    \end{tabular}
  \end{center}
\end{table}

\subsection{Corrections in the tunneling rate from Gaussian beam geometry}

In previous sections, we assume a separable confinement potential and tunneling only along the direction of gravity. However in the experimental system, the Gaussian geometry of the laser beams inevitably couple the axial and radial wave functions. This coupling   leads to  corrections in the  nearest-neighbor tunneling rate which now  depends on the thermal distribution of the radial modes. Notice that the Gaussian beam profile of a 1D lattice leads to the following trapping potential,
\begin{equation}
    V(X,Y,Z)=V_0-V_0\cos^2(k_L Z)\exp[-2(X^2+Y^2)/w_L^2],
\end{equation}
where $k_L=2\pi/\lambda_L$ is the lattice wave number, $w_L$ is the beam waist, and $V_0>0$ is the lattice depth. Expanding the trapping potential to second order of $X,Y$, we have
\begin{equation}
    V(X,Y,Z)\approx \bigg[V_0-\frac{1}{2}M\omega_R^2(X^2+Y^2)\bigg]\sin^2(k_L Z)+\frac{1}{2}M\omega_R^2(X^2+Y^2),
    \label{eq:rad}
\end{equation}
where the radial trapping frequency is given by $\omega_R=\sqrt{4V_0/Mw_L^2}$. Based on Eq.~(\ref{eq:rad}), an atomic gas with radial temperature $T_r$ feels an  effective lattice depth given by  $V_0-k_BT_r$. Although $k_BT_r\ll V_0$, it may still lead to  non-negligible corrections to the nearest-neighbor tunneling rate, which shows exponential dependence on lattice depth. Note that in the large-spin Hamiltonian discussed in previous sections, the interaction parameters are determined by thermal average over radial modes. To take into  account  the leading order effects of the thermal distribution, we replace the ground band tunnel coupling by 
\begin{equation}
    J_0(T_r)\approx \frac{4}{\sqrt{\pi}}E_{rec}^{1/4}(V_0-k_BT_r)^{3/4}\exp\bigg[-2\sqrt{\frac{V_0-k_BT_r}{E_{rec}}}\bigg].
\end{equation}
This correction leads to $\sim 40\%$ increase of nearest-neighbor $s$-wave interaction strength near the zero-crossing point.

\end{document}